\documentclass[11pt,fleqn]{article}

\usepackage[left=2.5cm,
            right=2.5cm,
            top=2.5cm,
            bottom=3.5cm,
            foot=1.5cm]{geometry}
 
\usepackage{graphicx}
\usepackage{orcidlink}
\usepackage{alltt}
\usepackage{lmodern}
\usepackage{amssymb, amsfonts, amsthm, mathtools}
\usepackage{latexsym, hyperref, url, moreverb}
\usepackage{enumerate}
\usepackage{xspace}
\usepackage{booktabs}
\usepackage{xcolor}
\usepackage[utf8]{inputenc}
\usepackage[T1]{fontenc}

\IfFileExists{upquote.sty}{\usepackage{upquote}}{}

\usepackage[]{natbib}
\graphicspath{{Figures/}}

\newenvironment{bsmatrix}{%
  \bigl[ \begin{smallmatrix} }{%
  \end{smallmatrix} \bigr]}

\newcommand{\Real}{\mathbb{R}}
\newcommand{\T}{{}^{\!\top}}
\newcommand{\x}{\boldsymbol{x}}
\newcommand{\Xspace}{\mathcal{X}}
\newcommand{\w}{\boldsymbol{w}}
\newcommand{\thetab}{\boldsymbol{\theta}}
\newcommand{\alphab}{\boldsymbol{\alpha}}
\newcommand{\Thetab}{\boldsymbol{\Theta}}
\newcommand{\pib}{\boldsymbol{\pi}}
\newcommand{\1}{\boldsymbol{1}}
\newcommand{\diff}{\text{d}}
\renewcommand{\bar}[1]{\overline{#1}}
\renewcommand{\hat}[1]{\widehat{#1}}
\renewcommand{\tilde}[1]{\widetilde{#1}}
\newcommand{\Multinomial}{\mathcal{M}}
\newcommand{\mub}{\boldsymbol{\mu}}
\newcommand{\Sigmab}{\boldsymbol{\Sigma}}
\newcommand{\Dirichlet}{\mathcal{D}}
\newcommand{\GammaFun}{\Gamma}
\newcommand{\GammaDistr}{\mathcal{G}}
\DeclareMathOperator{\Exp}{\mathbb{E}}
\DeclareMathOperator{\Median}{\text{Median}}

\begin{document}

\title{\LARGE\bfseries%
Assessing uncertainty in Gaussian\\ mixtures-based entropy estimation}

\author{
\large Luca Scrucca~\orcidlink{0000-0003-3826-0484}\\
\normalsize Department of Statistical Sciences, University of Bologna}

\date{\normalsize \today}

\maketitle

\begin{abstract}
\noindent%
Entropy estimation plays a crucial role in various fields, such as information theory, statistical data science, and machine learning.
However, traditional entropy estimation methods often struggle with complex data distributions. Mixture-based estimation of entropy has been recently proposed and gained attention due to its ease of use and accuracy.
This paper presents a novel approach to quantify the uncertainty associated with this mixture-based entropy estimation method using weighted likelihood bootstrap.
Unlike standard methods, our approach leverages the underlying mixture structure by assigning random weights to observations in a weighted likelihood bootstrap procedure, leading to more accurate uncertainty estimation.
The generation of weights is also investigated, leading to the proposal of using weights obtained from a specific Dirichlet distribution which, in conjunction with centered percentile intervals, yields the optimal setting to ensure empirical coverage closer to the nominal level.
Extensive simulation studies comparing different resampling strategies are presented and results discussed.
The proposed approach is illustrated by analyzing the log-returns of daily Gold prices at COMEX for the years 2014--2022, and the Net Rating scores, an advanced statistic used in basketball analytics, for NBA teams with reference to the 2022/23 regular season.
\end{abstract}

\clearpage

\tableofcontents	

\clearpage
\section{Introduction}

Entropy is a core concept in information theory \citep{Shannon:1948} providing a measure of uncertainty or randomness in a random variable \citep{Cover:Thomas:2006}.
The (differential) entropy of a continuous random variable $X \in \Real^d$ with probability density function (pdf) $f(\x)$ is defined as
\begin{equation*}
H(X) = - \Exp_X [\log f(X)] = - \int_{\Xspace} f(\x) \log f(\x) \,\diff \x ,
\end{equation*}
where $\Xspace = \{\x : f(\x) > 0 \}$ is the support of the random variable.

Closed-form expressions for the entropy of several univariate distributions are available in the literature \citep{Michalowicz:etal:2014}, but only few multivariate distributions admit a solution in analytic form.
For parametric pdfs $\{ f(\x; \thetab) : \thetab \in \Thetab \}$, when a closed-form expression is available, using the maximum likelihood estimate (MLE) $\hat{\thetab}$ guarantees that the estimate of the entropy is both asymptotically unbiased and efficient \citep[Theorem 7.3]{Kay:1993}.
If we cannot assume a specific parametric distribution, an estimate of the entropy is often obtained using nonparametric estimation of $f(\x)$ based on histogram, kernel density, or nearest neighbour estimators. This latter approach is, however, limited to low-dimensional cases, often $d \le 2$.
Recently, a simple and intuitive mixture-based estimator has been proposed by \citet{Robin:Scrucca:2023} using finite mixture models (FMMs).
The authors showed with extensive simulation studies the accuracy and efficiency of the proposal, which compares favorably against other estimators often used in practice.

The remainder of this paper is structured as follows:
Section~\ref{sec:mixent} reviews the mixture-based approach for entropy estimation and its advantages over traditional methods.
Section~\ref{sec:wlb} introduces our proposal for quantifying uncertainty in mixture-based entropy estimation. Specifically, we propose to adopt a weighted likelihood bootstrap (WLB) approach by assigning random weights drawn from a Dirichlet distribution when refitting mixture models.
Generation of weights is discussed in Section~\ref{sec:wlbwts}, where we introduce a novel proposal instead of the usually adopted uniform Dirichlet over the simplex.
Extensive simulation studies are presented in Section~\ref{sec:sim}, where the performance of different resampling strategies and percentile bootstrap intervals are compared.
Section~\ref{sec:applic} illustrates the methodology by analyzing entropy in two real-world datasets.
The final section concludes with a summary of findings and potential future research directions.

\section{Mixture-based entropy estimation}
\label{sec:mixent}

Assume that the probability density function of interest can be approximated by a finite mixture models (FMMs) with $G$ components, i.e.
\begin{equation}
f(\x) \approx f(\x; \thetab) = \sum_{k=1}^G \pi_k \, f_k(\x; \thetab_k),
\label{eq:fmm}
\end{equation}
where $\thetab = \{ \pib, \thetab_1, \dots, \thetab_G \}$ are the unknown parameters, with $\pib = (\pi_1, \dots, \pi_G)$ the mixing weights (s.t. $\pi_k > 0, \sum_{k=1}^G \pi_k = 1)$, and $\thetab_k$ the parameters associated to the density $f_k(\cdot)$ of the $k$-th component of the mixture ($k=1,\dots,G$).
Often, the densities $f_k(\cdot)$ are taken to belong to the same parametric family of distributions, but with different parameters $\thetab_k$.

\citet[][Lemma 11.2]{Wang:Madiman:2014} showed that the upper bound on the entropy $H(X)$ of the FMM distribution in \eqref{eq:fmm} is given by
\begin{equation*}
H(X) \le - \sum_{k=1}^G \pi_k \log\pi_k + \sum_{k=1}^G \pi_k \, H(f_k(\x; \thetab_k)).
\end{equation*}
Moreover, when the FMM is made up of Gaussian components, as introduced later in \eqref{eq:gmm}, \citet{Huber:etal:2008} showed that both a lower and an upper bounds can be obtained, respectively, as
\begin{align*}
H_\ell(X) & = - \sum_{k=1}^G \pi_k \log \left( \sum_{j=1}^G \pi_j \phi(\mub_k; \mub_j, \Sigmab_k + \Sigmab_j) \right), \\
H_u(X) & = - \sum_{k=1}^G \pi_k \log\pi_k + \sum_{k=1}^G \pi_k \left( \frac{d}{2} (1 + \log(2\pi)) + \frac{1}{2}\log|\Sigmab_k| \right).
\end{align*}
In practice, following a plug-in principle, lower and upper bound estimates are easily obtained by replacing unknown quantities with their estimates.

Recalling that, from a generative point of view, FMMs in \eqref{eq:fmm} can be re-expressed as a hierarchical model by introducing the multinomial discrete latent variable $Z = (Z_1, \dots, Z_G)\T$, where $Z_k = 1$ if the component of origin of $X$ in the mixture is equal to $k$ and 0 otherwise ($k = 1, \dots, G$), and assuming
\begin{align*}
Z & \sim \Multinomial(1, \pib) \\
X \mid (Z_k = 1) & \sim f_k(\x; \thetab_k),
\end{align*}
where $\Multinomial(1, \pib)$ is the multinomial distribution consisting of one draw on $G$ categories with probabilities $\pib = (\pi_1, \dots, \pi_G)\T$.
Then, by the law of total expectation, the entropy can be written as
\begin{align*}
H(X)
& = - \Exp_Z \left[ \Exp_{X \mid Z} \left[ \log f(X) \right]\right]
  = - \sum_{k=1}^G \pi_k \Exp_X \left[ \log f(X) \mid Z=k \right] \\
& = - \sum_{k=1}^G \pi_k \int_{\Xspace} f_k(\x; \thetab_k) \, \log f_k(\x; \thetab_k) \, \diff\x.
\end{align*}

Based on this formulation, \citet{Robin:Scrucca:2023} showed that a simple closed-formula mixture-based estimator for samples of size $n$ can be calculated as follows:
\begin{equation}
\hat{H}(X) = -\frac{1}{n} \sum_{i=1}^n \log \hat{f}(\x_i; \hat{\thetab}),
\label{eq:entropy}
\end{equation}
where $\hat{f}(\x_i; \hat{\thetab}) = \sum_{k=1}^G \hat{\pi}_k f_k(\x_i; \hat{\thetab}_k)$ is the mixture-based density estimate evaluated on the same data points used for fitting the model via EM algorithm \citep{Dempster:Laird:Rubin:1977, McLachlan:Krishnan:2008}.

Gaussian mixture models (GMMs) correspond to the case where the general FMM in \eqref{eq:fmm} is written as
\begin{equation}
f(\x; \thetab) = \sum_{k=1}^G \pi_k \, \phi( \x ; \mub_k, \Sigmab_k),
\label{eq:gmm}
\end{equation}
where $\thetab = \{\pib, \mub_1, \dots, \mub_G, \Sigmab_1, \dots, \Sigmab_G \}$ are the unknown parameters of the mixture, with $\pib = (\pi_1,\pi_2, \dots, \pi_G)$ the mixing weights, and $\phi(\x ; \mub_k, \Sigmab_k)$ the underlying multivariate Gaussian pdf of the $k$-th component with mean vector $\mub_k$ and covariance matrix $\Sigmab_k$.
The entropy estimate in \eqref{eq:entropy} can then be written as
\begin{align}
\hat{H}(X) &
= -\frac{1}{n} \sum_{i=1}^n \log
   \left\{
     \sum_{k=1}^G \hat{\pi}_k \, \phi(\x_i; \hat{\mub}_k, \hat{\Sigmab}_k)
   \right\},
\label{eq:entropy_mixgauss}
\end{align}
where $(\hat{\pi}_k, \hat{\mub}_k, \hat{\Sigmab}_k)$ are the estimated parameters of $k$-th component of the GMM obtained using the EM algorithm. For details see \citet[Sec. 2.2]{mclust:book:2023}.

The estimation procedure described above is available in the R package \texttt{mclustAddons} \citep{Rpkg:mclustAddons}.


\section{Uncertainty assessment via weighted likelihood bootstrap}
\label{sec:wlb}

Uncertainty assessment in entropy estimation is crucial for reliable interpretation of estimates.
Although some authors have addressed this issue in the context of entropy for discrete distributions \citep{Amiri:2014, Pesenti:etal:2017}, existing literature lacks methodologies specifically designed for differential entropy.
To fill this gap, we propose adopting a resampling approach based on the bootstrap \citep{Efron:Tibshirani:1993, Davison:Hinkley:1997} to quantify uncertainty effectively.

Standard nonparametric bootstrap (BS) is a resampling technique that relies on repeatedly drawing samples with replacement from the observed data \citep{Efron:1982}.
This procedure is equivalent to the repeated generation of random weights from a multinomial distribution with equal probabilities, i.e. $\Multinomial(1;\; \tfrac{1}{n}\1_n)$, where $\1_n$ is a vector of ones of length $n$.
In contrast, Weighted Likelihood Bootstrap (WLB) is a resampling procedure that generate bootstrap samples by assigning random weights to observations generated from a Dirichlet distribution over the $(n-1)$-simplex, i.e.
\begin{equation*}
(w_1,\dots,w_n) \sim \Dirichlet(\alphab),
\end{equation*}
where $\alphab = (\alpha_1, \dots, \alpha_n)\T$.
Therefore, the generated weights satisfy $0 \le w_i \le 1$ for all $i$ and $\sum_{i=1}^n w_i = 1$.

WLB was originally proposed by \citet{Newton:Raftery:1994} as a method for approximately sampling from a posterior distribution.
In its essence, WLB combines weighted likelihood estimation with bootstrap resampling using random weights as in the Bayesian bootstrap (BB) approach \citep{Rubin:1981}. Recently, Weighted Bayesian Bootstrap (WBB) has been proposed, enabling the inclusion of a prior distribution into statistical and machine learning models \citep{Newton:etal:2021}.

WLB requires the optimization of a weighted likelihood function, which for Gaussian mixtures is given by
\begin{equation*}
\ell^W(\thetab) =
\sum_{i=1}^n w_i
\log\left\{
            \sum_{k=1}^G \pi_k \, \phi( \x_i ; \mub_k, \Sigmab_k)
    \right\}.
\end{equation*}
EM algorithm can be used for estimation by defining the weighted complete-data log-likelihood:
\begin{equation}
\ell_C^W(\thetab) =
\sum_{i=1}^n \sum_{k=1}^G z_{ik} \, w_i
\log\left\{
            \pi_k \, \phi( \x_i ; \mub_k, \Sigmab_k)
    \right\},
\label{eq:wcomploglik}
\end{equation}
where 
$\thetab = (\pi_1, \dots, \pi_{G-1}, \mub_1, \dots, \mub_G, \Sigmab_1, \dots, \Sigmab_G)$ 
are the parameters to be estimated, and 
$z_{ik} = \pi_k \phi( \x_i ; \mub_k, \Sigmab_k) / \sum_{g = 1}^G \pi_g \phi( \x_i ; \mub_g, \Sigmab_g)$, 
the conditional probability that the $i$-th observation arose from mixture component $k$.

Let GMM$(M, G)$ be the Gaussian mixture model $M$ defined by a parsimonious covariance decomposition \citep[Sec. 2.2.1]{mclust:book:2023} with $G$ number of mixture components. These represent tuning parameters that can be set by the researcher or determined using a model selection criterion, such as the Bayesian information criterion \citep[BIC;][]{Schwarz:1978} or the integrated complete-data likelihood criterion \citep[ICL;][]{Biernacki:Celeux:Govaert:2000}.

\medskip

The proposed WLB procedure for a GMM$(M, G)$ can be described as follows:

\begin{enumerate}

\item generate weights as $(w_1,\dots,w_n) \sim \Dirichlet(\alphab)$; \smallskip

\item fit GMM$(M, G)$ using the data $\{\x_i, w_i \}_{i=1}^n$ by maximizing \eqref{eq:wcomploglik} to get the bootstrap estimate of parameters
$\hat{\thetab}^{\,*}_b =
\left\{
  \hat{\pi}^{\,*}_k, \hat{\mub}^{\,*}_k, \hat{\Sigmab}^{\,*}_k
\right\}_{k=1}^{G}$;

\item compute the bootstrap estimate of entropy as
\begin{equation*}
\hat{H}^*_b(X) = -\frac{1}{n\bar{w}} \sum_{i=1}^n w_i \log
                  \left\{ \sum_{k=1}^G \hat{\pi}^{\,*}_k \phi(\x_i; \hat{\mub}^{\,*}_k, \hat{\Sigmab}^{\,*}_k)
                  \right\},
\end{equation*}
where $\bar{w} = n^{-1} \sum_{i=1}^n w_i$;

\item replicate steps 1--3 a large number of times, say $B$, to obtain the WLB distribution $\hat{H}^*(X) = \left\{ \hat{H}^*_b(X) \right\}_{b=1}^B$ as an approximate posterior sampling.
\end{enumerate}

Once the bootstrap distribution is obtained, various measures can be derived to assess the GMM-based estimate $\hat{H}(X)$ of entropy.
For instance, an estimate of the bias can be computed as
\begin{equation*}
\text{Bias}(\hat{H}(X)) = \bar{\hat{H}^*}(X) - \hat{H}(X),
\end{equation*}
where $\bar{\hat{H}^*}(X) = B^{-1} \sum_{b=1}^B \hat{H}^*_b(X)$, while a bootstrap estimate of the standard error is obtained as
\begin{equation*}
\text{SE}(\hat{H}(X)) = \sqrt{ \frac{1}{B-1} \sum_{b=1}^B \left( \hat{H}^*_b(X) - \bar{\hat{H}^*}(X) \right)^2 }.
\end{equation*}

WLB intervals at the level $(1-\alpha)100\%$ can be derived using the \emph{bootstrap percentile} method, i.e. by computing
\begin{equation}
\left[ \hat{H}^*(X; \alpha/2), \; \hat{H}^*(X; 1-\alpha/2) \right],
\label{eq:percint}
\end{equation}
where $\hat{H}^*(X; q)$ is the 100$q$-th percentile of the WLB distribution.
Percentile intervals are straightforward to compute and have the advantage of being range-preserving, ensuring the interval lies within the proper bounds of the parameter space.
However, despite their simplicity and intuitive reasonableness, percentile intervals have been criticized because they approximate the sampling distribution of $\hat{H}(X) - H(X)$ with $\hat{H}(X) - \hat{H}^*(X)$ and not with $\hat{H}^*(X) - \hat{H}(X)$, which is what the bootstrap prescribes \citep{Hall:1988}.


According to the bootstrap principle, we should aim at computing an interval such that
\begin{gather*}
\Pr\left( \hat{H}^*(X;\alpha/2) - \hat{H}(X) \le \hat{H}(X) - H(X) \le \hat{H}^*(X; 1-\alpha/2) - \hat{H}(X) \right) \approx 1-\alpha \\
\Pr\left( 2\hat{H}(X) - \hat{H}^*(X; 1-\alpha/2) \le H(X) \le 2\hat{H}(X) - \hat{H}^*(X; \alpha/2) \right) \approx 1-\alpha.
\end{gather*}
Notice that, if $\hat{H}^*_b(X) - \hat{H}(X)$ is the bias for the $b$-th bootstrap sample, then a bias-corrected estimate is given by
\begin{equation*}
\hat{H}^{*c}_b(X) = \hat{H}(X) - (\hat{H}^*_b(X) - \hat{H}(X)) = 2\hat{H}(X) - \hat{H}^*_b(X).
\end{equation*}
The aforementioned arguments result in a simple bias-corrected version known as the \emph{centered bootstrap percentile} method (\citealp[Sec. 3.3]{Manly:2006}; \citealp{Singh:Xie:2010}) or \emph{basic bootstrap intervals} \citep[Sec. 5.2]{Davison:Hinkley:1997}.
Given the differing roles of the bootstrap distribution in the two percentile methods, the resulting intervals will not be equal, especially when the bootstrap distribution is skewed or biased.

Thus, a WLB centered percentile interval at the level $(1-\alpha)100\%$ can be computed using the percentiles from the distribution of bias-corrected entropy estimates $\hat{H}^{*c}(X) = \left\{ \hat{H}^{*c}_b(X) \right\}_{b=1}^B$, i.e. by computing
\begin{equation}
\left[ \hat{H}^{*c}(X; \alpha/2), \; \hat{H}^{*c}(X; 1-\alpha/2) \right] =
\left[ 2\hat{H}(X) - \hat{H}^*(X; 1-\alpha/2), \; 2\hat{H}(X) + \hat{H}^*(X; \alpha/2) \right],
\label{eq:cpercint}
\end{equation}
where $\hat{H}^{*c}(X; q)$ is the 100$q$-th percentile of the bias-corrected WLB distribution.

\section{On the generation of weights in WLB}
\label{sec:wlbwts}

In WLB, as well as in BB and WBB methods, the weights associated with each observation are generated from a Dirichlet distribution, i.e. $\w \sim \Dirichlet(\alphab)$ with $\alphab = \alpha\1_n$.
In general, generating random values from a $\Dirichlet(\alphab)$ distribution is equivalent to generating them from independent Gamma distributions $\GammaDistr(\alpha, \beta = 1)$, resulting in both the expected value and the variance being equal to $\alpha$. It is worth noting that when $\alpha = 1$, an Exponential distribution with a rate parameter of 1 is obtained.

Weights generated independently according to $w_i \sim \GammaDistr(\alpha, \beta = 1)$, for $i=1,\dots,n$, can be rescaled in various ways.
Often, weights are scaled such that they sum to one or, more conveniently, such that their sum is equal to the number of observations $n$.
In the latter case, the rescaling is obtained as $\tilde{w}_i = w_i/\bar{w}$, where $\bar{w} = n^{-1} \sum_{i=1}^n w_i$, so $\sum_{i=1}^n \tilde{w}_i = n$. Using this rescaling, weights can be seen as the contribution of each observation to the overall sample information.
Note that in the unweighted case, each observation brings a unit contribution. In contrast, in the nonparametric bootstrap, an observation contributes 0 if it is never sampled, 1 if it is sampled once, 2 if it is sampled twice, and so forth.

As previously mentioned, a typical choice is $\alphab = (1,1,\dots,1)\T = \1_n$, which results in random weights being generated uniformly over the simplex.
However, this may not always represent the optimal setting, and few studies have been dedicated to exploring this aspect. A notable exception is the recommendation by \citet[Sec. 10.2.1]{Shao:Tu:1995} to set $\alphab = (4,4,\dots,4)\T = 4\1_n$.

Here, we propose using $\alpha = 0.8137$, a value derived by mimicking a well-known result in standard nonparametric bootstrap. Recalling that the probability of an observation being included at least once in a nonparametric bootstrap sample is $1 - \exp(-1) \approx 0.6321$ as $n \to \infty$ \citep[Problem 17.7]{Efron:Tibshirani:1993}, we aim to determine the value of $\alpha$ that ensures, on average, a median weight of $0.6321$ is assigned to each observation.

Suppose weights are generated as $\left\{ w_i \sim \GammaDistr(\alpha, \beta = 1) \right\}_{i=1}^n$, thus $\Exp(w_i) = \alpha$, and rescale the weights by their average to obtain $\tilde{w}_i = w_i/\alpha$.
We aim at finding the value of $\alpha$ such that $\Exp[\Median(\tilde{w})] = 0.6321$, i.e the value at which, on average, the median weight assigned to an observation equals the probability of inclusion in standard nonparametric bootstrap.
The use of the median is justified as a robust measure of centrality, which is required due to the highly skewed distributions involved.

\begin{figure}
\centering
\includegraphics[width=0.8\textwidth]{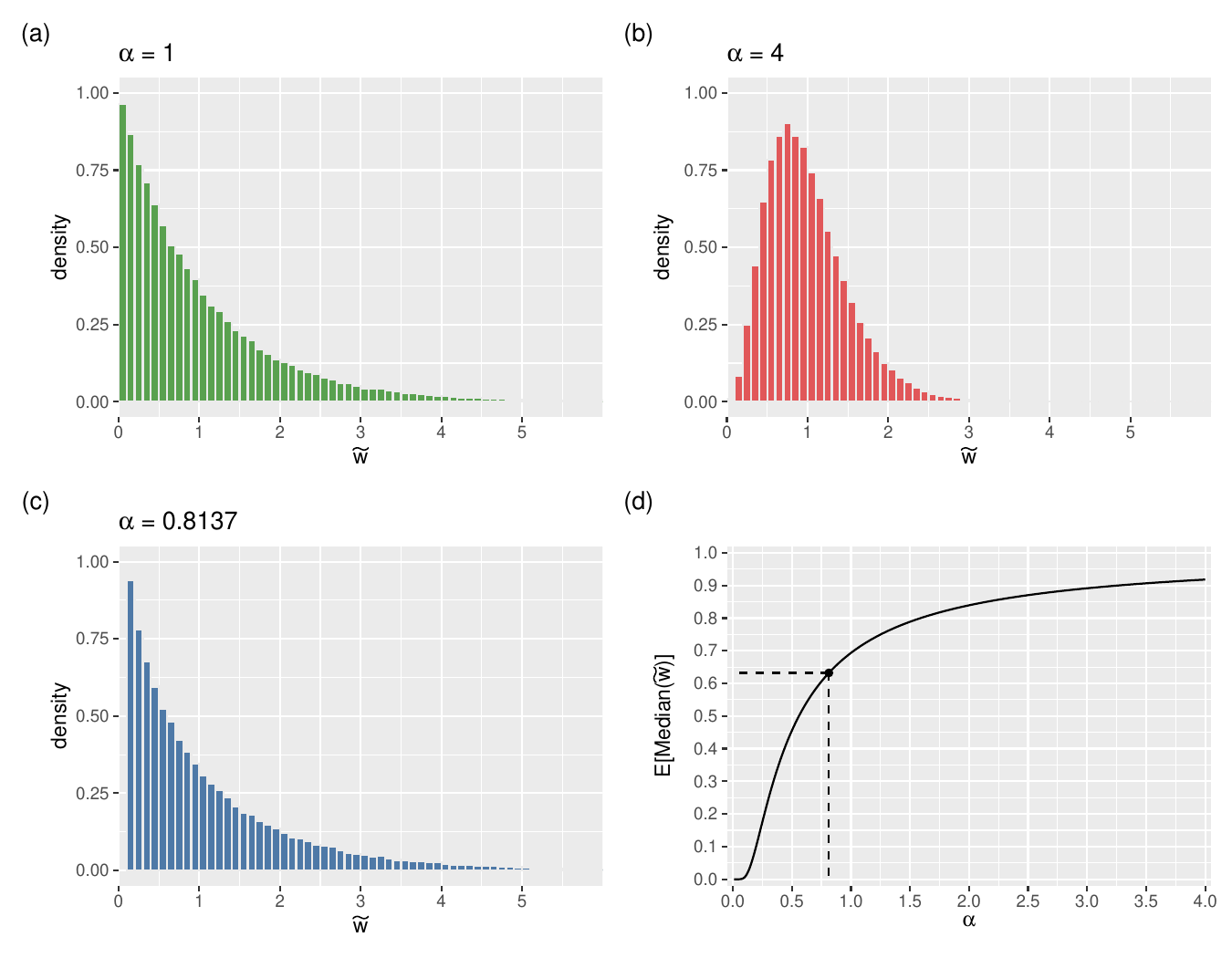}
\caption{Panels (a)-(c) show the distribution of scaled weights $\tilde{w}_i$ obtained by varying the parameter $\alphab = \alpha\1_n$ of a Dirichlet distribution. Panel (d) displays the curve of $\Exp[\Median(\tilde{w})]$ as a function of $\alpha$, with the highlighted value $\alpha = 0.8137$ indicating the parameter value that ensures, on average, the median weight assigned to an observation matches the probability of inclusion in standard nonparametric bootstrap.}
\label{fig:alpha_selection}
\end{figure}

Panels (a) and (b) of Figure~\ref{fig:alpha_selection} show the distribution of scaled weights $\tilde{w}_i$ corresponding to the typical value of $\alpha=1$ and $\alpha=4$, the latter being the value suggested by \citet[Sec. 10.2.1]{Shao:Tu:1995}.
In the former case, the distribution exhibits significant skewness, with most data points having small weights and only a few receiving larger weights.
Conversely, in the latter case, the weights tend to be more similar, thereby reducing diversity across resamples.
Figure~\ref{fig:alpha_selection}d displays the curve of $\Exp[\Median(\tilde{w})]$ as a function of $\alpha$, revealing the sought value of $\alpha = 0.8137$ obtained by root finding. Figure~\ref{fig:alpha_selection}c illustrates the distribution of weights corresponding to the identified value of $\alpha = 0.8137$ for the random generation of weights in a WLB procedure. The distribution of scaled weights exhibits larger skewness compared to the standard case, thereby increasing the variability of the generated weights.
Contrary to the suggestion by \citet{Shao:Tu:1995}, the value of $\alpha$ is reduced, thereby increasing the diversity of generated resamples. As will be shown in Section~\ref{sec:sim}, this has beneficial effects on the empirical coverage of credible intervals for the entropy.

\clearpage
\section{Simulation studies}
\label{sec:sim}

In this section, we present the results of a series of comprehensive simulation studies designed to examine the performance of different WLB strategies and to compare them with standard nonparametric and parametric bootstrap approaches, as described in Section~\ref{sec:simmeth}.
The simulation settings examined coincide in part with those analyzed in \citet{Robin:Scrucca:2023}, and are described in detail in Section~\ref{sec:simset}.
A summary of the main findings of the simulation studies is presented and discussed in Section~\ref{sec:simres}.

\subsection{Resampling methods under comparison}
\label{sec:simmeth}

The bootstrap approaches we aim at comparing are the following:

\begin{itemize}

\item nonparametric bootstrap (BS), with samples obtained by resampling with replacement from and with the same size as the original data;

\item parametric bootstrap (PB), with samples obtained by generating synthetic data of the same size from the fitted GMM on the original data;

\item weighted likelihood bootstrap (WLB) with samples obtained by assigning to the original simulated data a set of weights generated from a Dirichlet distribution with parameter $\alphab = \alpha\1_n$, where $\alpha = \{1, 4, 0.8137\}$.

\end{itemize}

\subsection{Data generation settings}
\label{sec:simset}

The distributions investigated in the simulation study are the following:

\begin{itemize}

\item Gaussian distribution with parameters $\mu = 0$ and $\sigma = 1$, whose entropy is
\begin{equation*}
H(X) = \frac{1}{2} (1 + \log(2\pi) + \log(\sigma^2)) \approx 1.4189.
\end{equation*}

\item $t$ distribution with $\text{df} = 3$ degrees of freedom, whose entropy is
\begin{equation*}
H(X) =
\frac{\text{df}+1}{2} \left[ \psi\left( \frac{\text{df}+1}{2} \right) - \psi\left( \frac{\text{df}}{2} \right) \right] +
\log \left[ \sqrt{\text{df}}\; \text{B}\left(\frac{\text{df}}{2},\frac{1}{2} \right)\right] \approx 1.7735,
\end{equation*}
where $\psi()$ is the digamma function, and $\text{B}()$ the beta function.

\item Mixed-Gaussian distribution with parameters $\mu=2$ and $\sigma=1$; the distribution is bimodal with modes at $\pm 2$ and entropy \citep{Michalowicz:etal:2014}
\begin{equation*}
H(X) = \frac{1}{2} (1 + \log(2\pi) + \log(\sigma^2)) + 0.633 \approx 2.0519.
\end{equation*}

\item Laplace distribution with parameters $\mu = 0$ and $\beta = 2$, whose entropy is
\begin{equation*}
H(X) = 1 + \log(2\beta) \approx 2.3863.
\end{equation*}

\item Bivariate ($d=2$) Gaussian distribution with parameters $\mu = [0,0]\T$ and ${\Sigmab = \begin{bsmatrix} 1.0 & 0.8 \\ 0.8 & 2.0 \end{bsmatrix}}$, whose entropy is given by
\begin{equation*}
H(X) = \frac{d}{2} (1 + \log(2\pi)) + \frac{1}{2} \log|\Sigmab| \approx 2.9916.
\end{equation*}

\item Multivariate ($d=10$) independent $\chi^2$ distributions with $\text{df}=5$ degrees of freedom for each dimension, thus having entropy equal to
\begin{equation*}
H(X) = \sum_{j = 1}^{10} H(X_j) = d \left[ \log(2) + \log\GammaFun(\text{df}/2) + \text{df}/2 + (1-\text{df}/2) \psi(\text{df}/2) \right] \approx 24.2309,
\end{equation*}
where $\GammaFun()$ is the gamma function, and $\psi()$ the digamma function.

\end{itemize}

Data were simulated from the aforementioned distributions for sample sizes $n = \{100, 200, 500, 1000 \}$. For each setting, 1,000 replications were generated.
A GMM was then fitted to each synthetic dataset using the \texttt{mclust} R package \citep{Rpkg:mclust}, with number of mixture components and parsimonious variances (or covariance matrices in case of multidimensional data) selected using the Bayesian information criterion \citep[BIC;][]{Schwarz:1978}.

\subsection{Results}
\label{sec:simres}

Tables \ref{tab:t_df=3} and \ref{tab:Gaussian_mu=0_sigma=1}--\ref{tab:ChiSq_df=5_d=10} in Appendix~\ref{ap:simres} present, for each investigated distribution, summaries of the simulation results for the bootstrap methods under comparison.
For the same settings, the associated Figures \ref{fig:t_df=3} and \ref{fig:Gaussian_mu=0_sigma=1}--\ref{fig:ChiSq_df=5_d=10} in Appendix~\ref{ap:simres} show the empirical coverage of different bootstrap intervals (panel a) and the ratio of bootstrap absolute bias over the bootstrap standard error (panel b).

Across all scenarios examined in our simulation studies, the nonparametric bootstrap, the parametric bootstrap, and the WLB with $\alpha=1$ exhibit remarkably similar coverage behavior. As expected, empirical coverage improves with larger sample sizes.
Conversely, the WLB procedure with $\alpha=0.8137$ provides empirical coverage values that are always closer to the nominal level, whereas the WLB with $\alpha=4$ consistently produces coverage levels significantly lower than the nominal level across all tested settings.
Moreover, centered bootstrap percentile intervals outperformed standard percentile bootstrap intervals in terms of empirical coverage. In fact, centered percentile intervals demonstrated empirical coverage levels consistently meeting or exceeding the nominal level.
Finally, when deciding between bootstrap percentile interval types, the absolute bias-to-standard error ratio provides valuable insight. A ratio exceeding 0.2 indicates a preference for centered percentile intervals, as these intervals tend to achieve empirical coverage levels equal to or greater than the nominal level, although with slightly wider intervals (not shown here).

To summarize the main results of our investigation, the WLB with $\alpha=0.8137$ emerges as the recommended method for assessing uncertainty in GMM-based entropy estimation. At the same time, the adoption of centered percentile intervals seems to offer the most prudent choice to ensure empirical coverage close to the nominal level.

\begin{table}
\caption{Simulation results for data generated from a $t$ distribution with $\text{df} = 3$ degrees of freedom. For each bootstrap method and sample size, the table reports averaged values of the entropy estimates, bootstrap estimates of bias, standard error, and percentile interval limits, along with their empirical coverage.}
\label{tab:t_df=3}
\small
\centering
\addtolength{\tabcolsep}{-0.6ex}
\begin{tabular}{lccccccccc}
\toprule
 & & & & \multicolumn{3}{c}{95\% percentile interval} & \multicolumn{3}{c}{95\% centered perc. interval} \\
\cmidrule(l{3pt}r{3pt}){5-7} \cmidrule(l{3pt}r{3pt}){8-10}
Sample size & \textbf{Estimate} & \textbf{Bias} & \textbf{SE} & lower & upper & \textbf{Coverage} & lower & upper & \textbf{Coverage}\\
\midrule
\multicolumn{10}{l}{Nonparametric bootstrap}\\[1ex]
 100 & 1.7565 & 0.0378 & 0.1160 & 1.4725 & 1.9256 & 0.890 & 1.5877 & 2.0413 & 0.927\\
 200 & 1.7621 & 0.0193 & 0.0775 & 1.5866 & 1.8890 & 0.921 & 1.6351 & 1.9370 & 0.939\\
 500 & 1.7703 & 0.0092 & 0.0488 & 1.6651 & 1.8554 & 0.936 & 1.6851 & 1.8754 & 0.953\\
1000 & 1.7734 & 0.0047 & 0.0344 & 1.7015 & 1.8355 & 0.933 & 1.7112 & 1.8452 & 0.933\\
\midrule
\multicolumn{10}{l}{Parametric bootstrap}\\[1ex]
 100 & 1.7565 & 0.0238 & 0.0991 & 1.5311 & 1.9181 & 0.859 & 1.5945 & 1.9818 & 0.908\\
 200 & 1.7621 & 0.0105 & 0.0727 & 1.6073 & 1.8909 & 0.899 & 1.6332 & 1.9168 & 0.926\\
 500 & 1.7703 & 0.0032 & 0.0472 & 1.6757 & 1.8599 & 0.948 & 1.6809 & 1.8647 & 0.949\\
1000 & 1.7734 & 0.0020 & 0.0336 & 1.7086 & 1.8396 & 0.926 & 1.7072 & 1.8384 & 0.925\\
\midrule
\multicolumn{10}{l}{Weighted likelihood bootstrap ($\alpha = 1$)}\\[1ex]
 100 & 1.7565 & 0.0261 & 0.1013 & 1.5349 & 1.9286 & 0.890 & 1.5837 & 1.9784 & 0.914\\
 200 & 1.7621 & 0.0151 & 0.0723 & 1.6092 & 1.8914 & 0.924 & 1.6333 & 1.9154 & 0.927\\
 500 & 1.7703 & 0.0073 & 0.0471 & 1.6730 & 1.8568 & 0.941 & 1.6836 & 1.8675 & 0.947\\
1000 & 1.7734 & 0.0041 & 0.0337 & 1.7047 & 1.8362 & 0.928 & 1.7105 & 1.8421 & 0.925\\
\midrule
\multicolumn{10}{l}{Weighted likelihood bootstrap ($\alpha = 4$)}\\[1ex]
 100 & 1.7565 & 0.0063 & 0.0507 & 1.6523 & 1.8499 & 0.630 & 1.6632 & 1.8611 & 0.621\\
 200 & 1.7621 & 0.0039 & 0.0366 & 1.6878 & 1.8305 & 0.629 & 1.6937 & 1.8364 & 0.633\\
 500 & 1.7703 & 0.0019 & 0.0238 & 1.7224 & 1.8153 & 0.657 & 1.7252 & 1.8181 & 0.658\\
1000 & 1.7734 & 0.0011 & 0.0170 & 1.7394 & 1.8058 & 0.662 & 1.7410 & 1.8073 & 0.663\\
\midrule
\multicolumn{10}{l}{Weighted likelihood bootstrap ($\alpha = 0.8137$)}\\[1ex]
 100 & 1.7565 & 0.0319 & 0.1124 & 1.5080 & 1.9437 & 0.917 & 1.5690 & 2.0056 & 0.939\\
 200 & 1.7621 & 0.0184 & 0.0798 & 1.5916 & 1.9028 & 0.938 & 1.6212 & 1.9324 & 0.951\\
 500 & 1.7703 & 0.0088 & 0.0520 & 1.6622 & 1.8654 & 0.963 & 1.6753 & 1.8782 & 0.966\\
1000 & 1.7734 & 0.0048 & 0.0373 & 1.6971 & 1.8427 & 0.954 & 1.7042 & 1.8495 & 0.953\\
\bottomrule
\end{tabular}
\end{table}

\begin{figure}
\centering
\includegraphics[width=\textwidth]{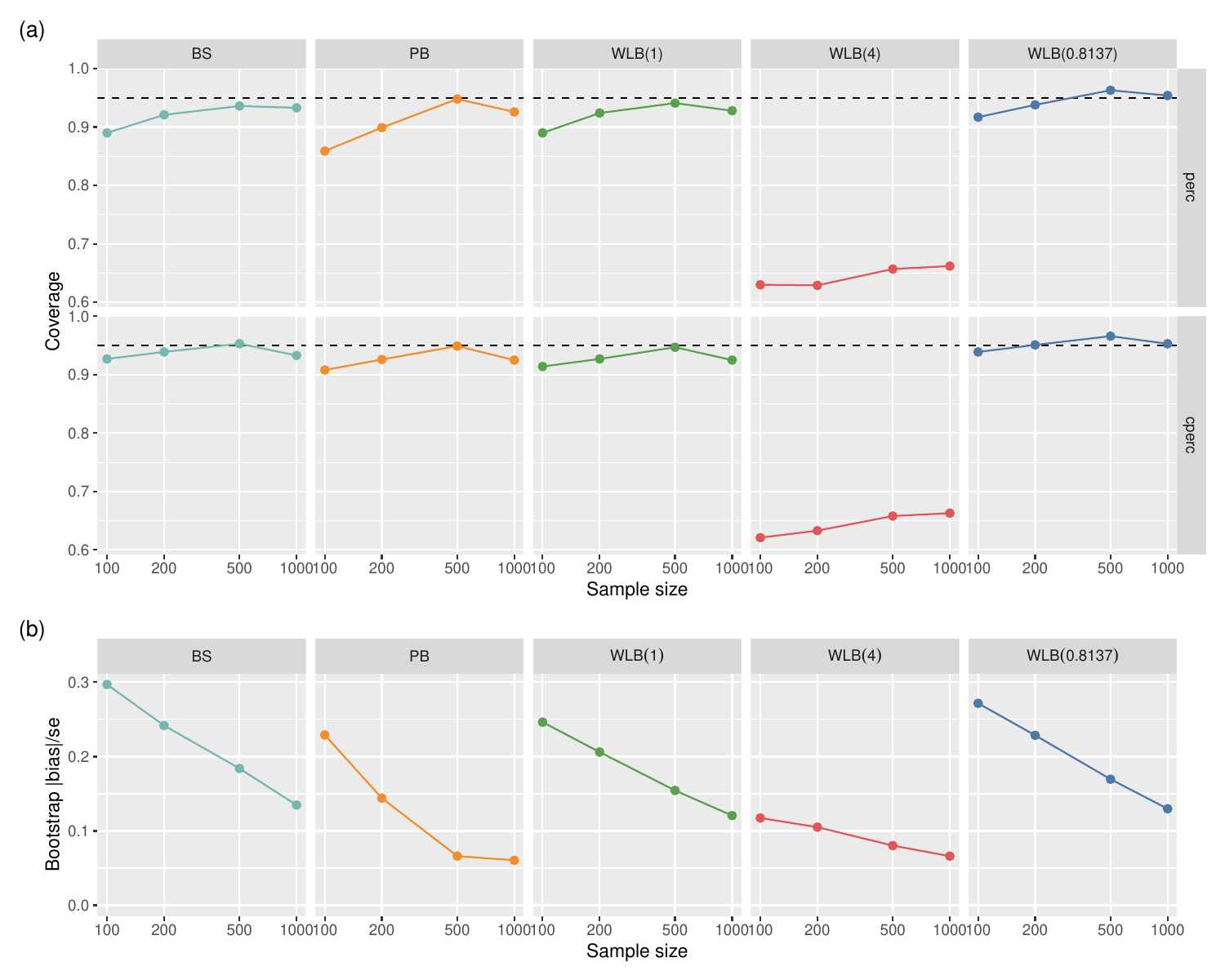}
\caption{Simulation results for data generated from a $t$ distribution with $\text{df} = 3$ degrees of freedom.
Panel (a) shows the empirical coverage for 1000 replications as function of the sample size for different bootstrap methods (BS = nonparametric; PB = parametric; WLB = weighted likelihood bootstrap with weights generated from a Dirichlet with different values of $\alpha$) and two type of bootstrap intervals (perc = percentile; cperc = centered percentile).
Panel (b) show the ratio of bias (in absolute value) over the standard error for each bootstrap procedure as function of sample size.}
\label{fig:t_df=3}
\end{figure}
\clearpage

\clearpage
\section{Data applications}
\label{sec:applic}

\subsection{Gold prices data}

Gold is often seen as a secure asset during periods of political and economic instability.
While the price of gold has generally trended upwards over the long term, short-term fluctuations of gold price can be substantial.
Given its propensity to appreciate along with the general increase in prices of other goods and services, it is considered a valuable safeguard against inflation.
For these reasons, gold is often regarded as a reliable investment that can protect investors' wealth in times of economic turmoil.

Here we analyze the differential entropy of the daily log returns of gold prices from 2014 to 2022.
By computing and comparing the differential entropy for each year, the relative uncertainty in gold prices over time can be assessed. Higher entropy values indicate greater uncertainty, suggesting more variability and potentially more unpredictable price movements. Conversely, lower entropy values imply lower uncertainty and more stable price behavior.

\begin{figure}
\centering
\includegraphics[width=0.8\textwidth]{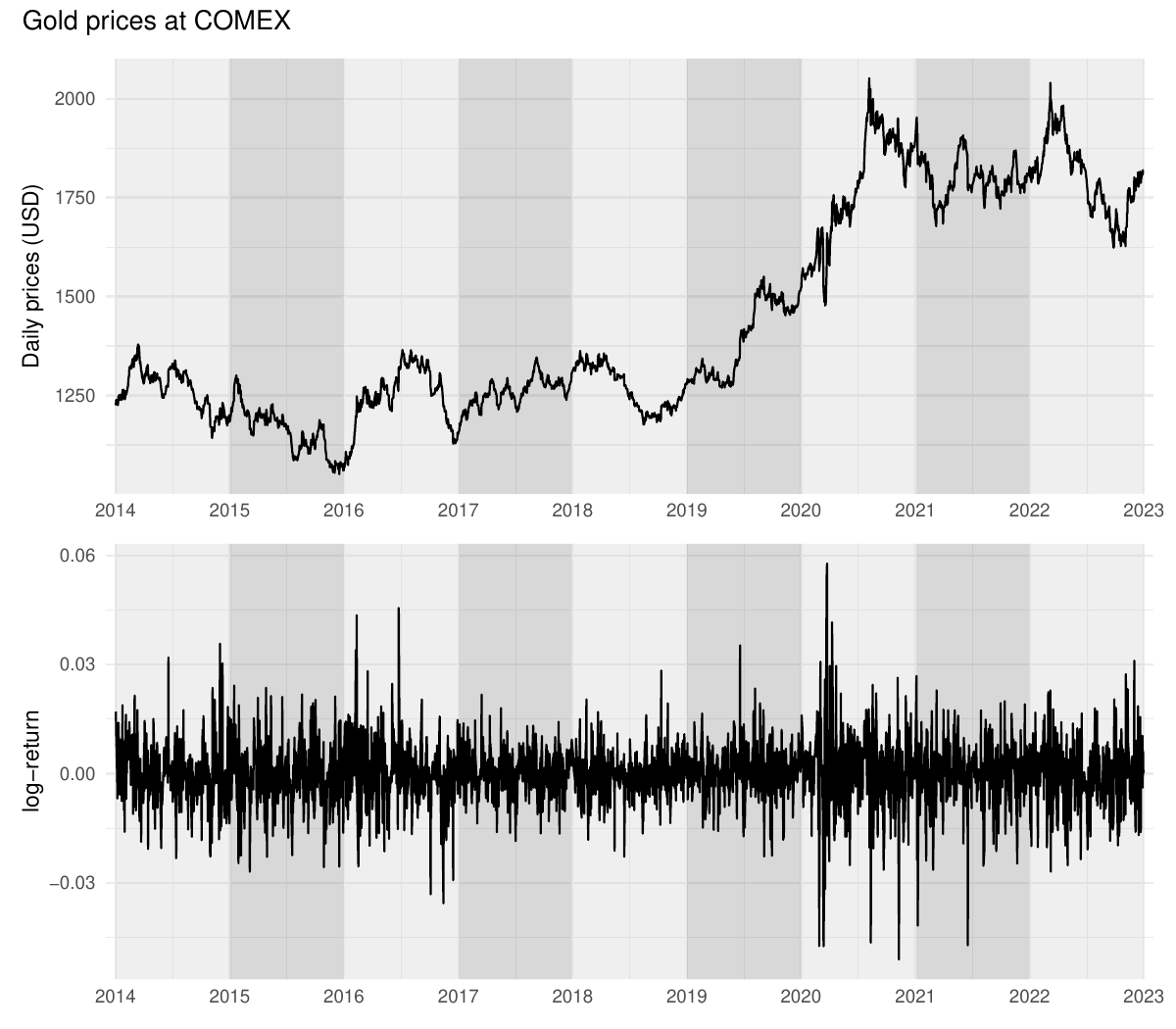}
\caption{Daily gold prices at COMEX for the years 2014--2022 (top panel) and the corresponding log-returns over the same period (bottom panel).}
\label{fig1:gold_prices}
\end{figure}

Figure~\ref{fig1:gold_prices} shows the daily Gold prices (per troy ounce) at Commodity Exchange Inc. (COMEX), the primary futures and options market for trading metals such as gold, silver, copper, and aluminum.
The daily gold prices shown on the top panel exhibit significant fluctuations throughout the period. Starting from around \$1,300 in 2014, the prices gradually increased, reaching a peak of around \$1,900 in 2020. After this peak, the prices experienced a decline but remained above \$1,500 for most of the remaining period until the end of 2022.
The log-returns shown in the bottom panel of Figure~\ref{fig1:gold_prices} display a high degree of volatility, with frequent fluctuations above and below zero, suggesting significant price movements in either direction throughout the period.

\begin{figure}
\centering
\includegraphics[width=0.8\textwidth]{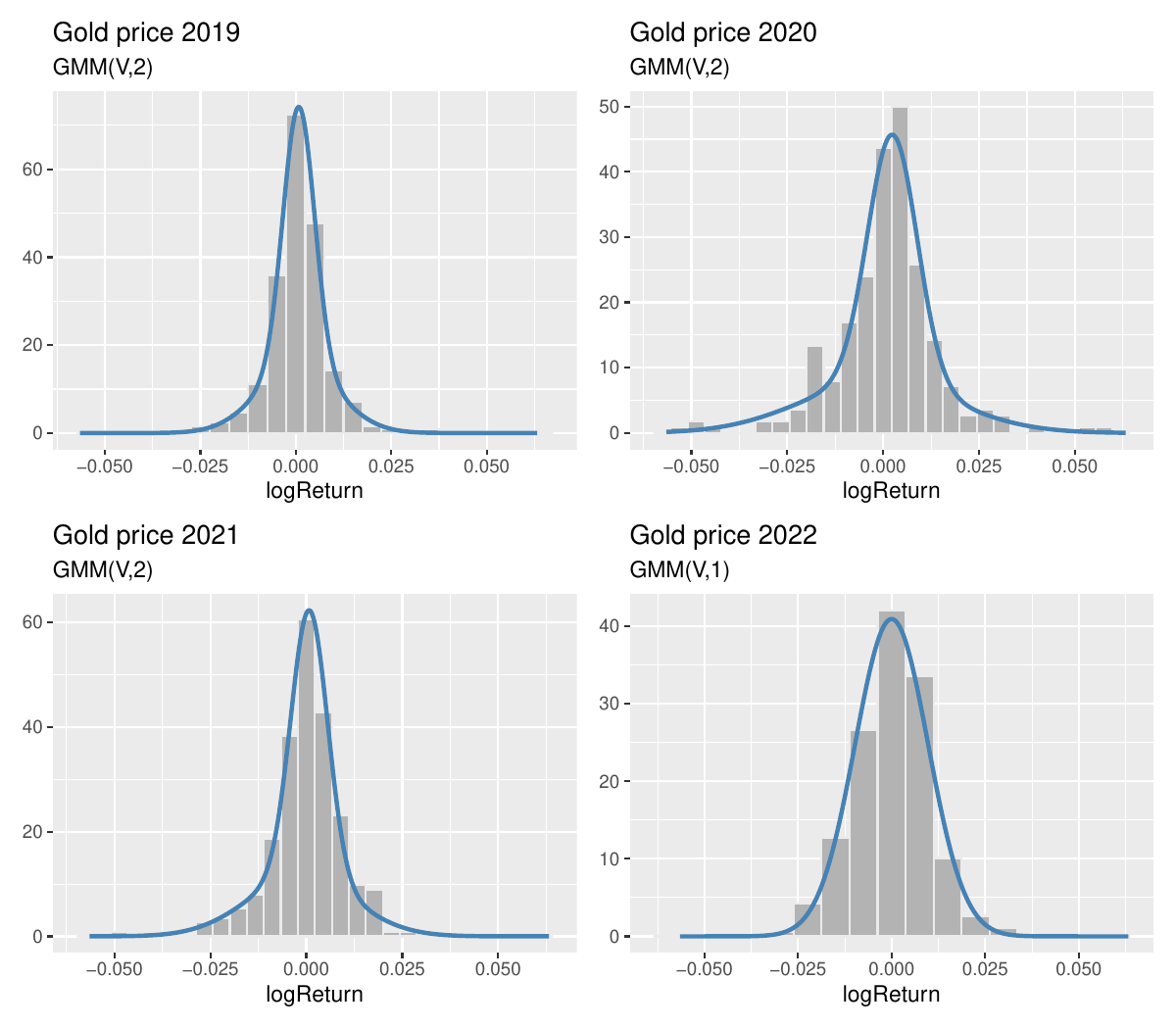}
\caption{Distribution of gold price log-returns for each year from 2019 to 2022, with histograms representing the empirical distributions and lines indicating the density estimates obtained by fitting Gaussian Mixture Models (GMMs) to the observed data.}
\label{fig2:gold_prices}
\end{figure}

To investigate the volatility in the gold market during the selected time frame we fitted Gaussian mixtures to the log-return daily prices for each year.
Figure~\ref{fig2:gold_prices} shows the histograms and fitted GMM densities for the years 2019--2022.
The volatility in the gold market, as captured by the distribution of daily log-returns, varied significantly over these four years, with 2020 exhibiting a higher volatility characterized by several larger positive and negative returns, while year 2019, and to a lesser extend 2021 and 2022, showed relatively lower volatility.

\begin{figure}
\centering
\includegraphics[width=0.8\textwidth]{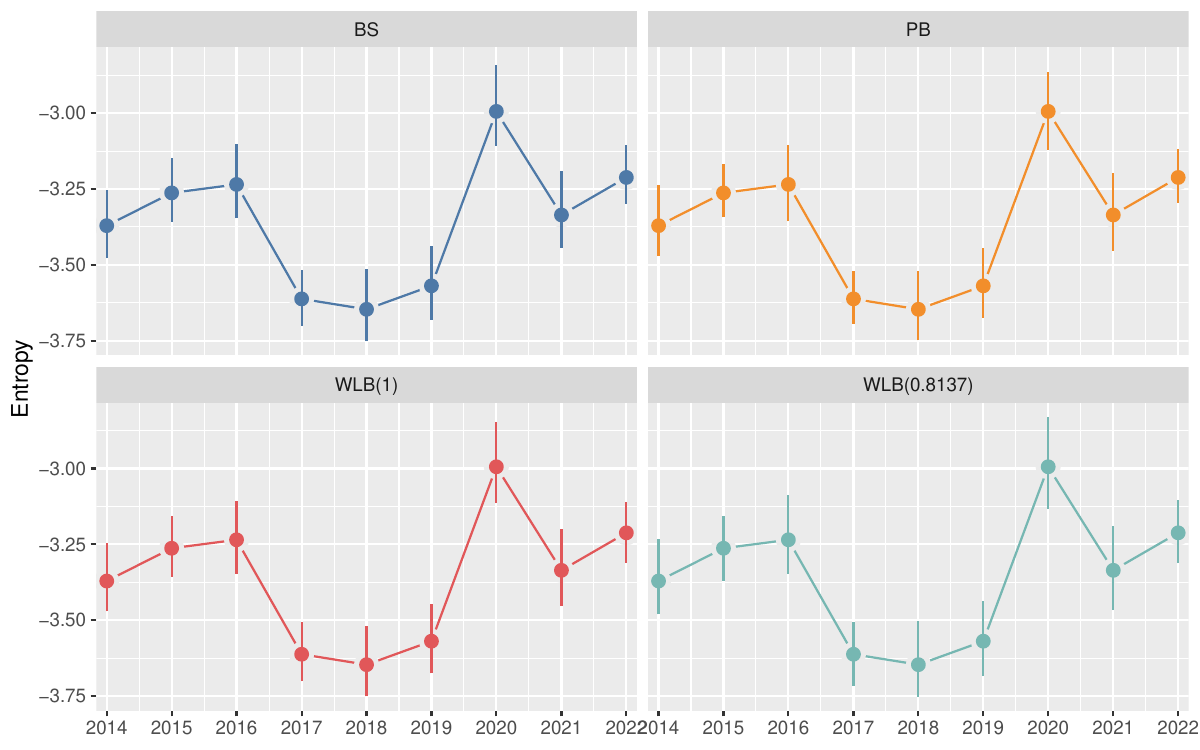}
\caption{Gaussian mixtures-based entropy estimates for each year from 2014 to 2022, with the associated 95\% centered percentile intervals obtained using various bootstrap methods (BS = nonparametric; PB = parametric; WLB = weighted likelihood bootstrap with weights generated from a Dirichlet with $\alpha=1$ and $\alpha=0.8137$).}
\label{fig3:gold_prices}
\end{figure}

Figure~\ref{fig3:gold_prices} displays the trend of GMM-based entropy estimates from 2014 to 2022, along with their associated 95\% centered percentile intervals obtained using various bootstrap procedures.
Despite being computed using different bootstrap techniques, the intervals represented by the vertical bars appear remarkably similar over the entire time period. This suggests that in this case the uncertainty associated with the entropy estimates is consistent, regardless of the specific bootstrap method employed.

\begin{figure}
\centering
\includegraphics[width=0.8\textwidth]{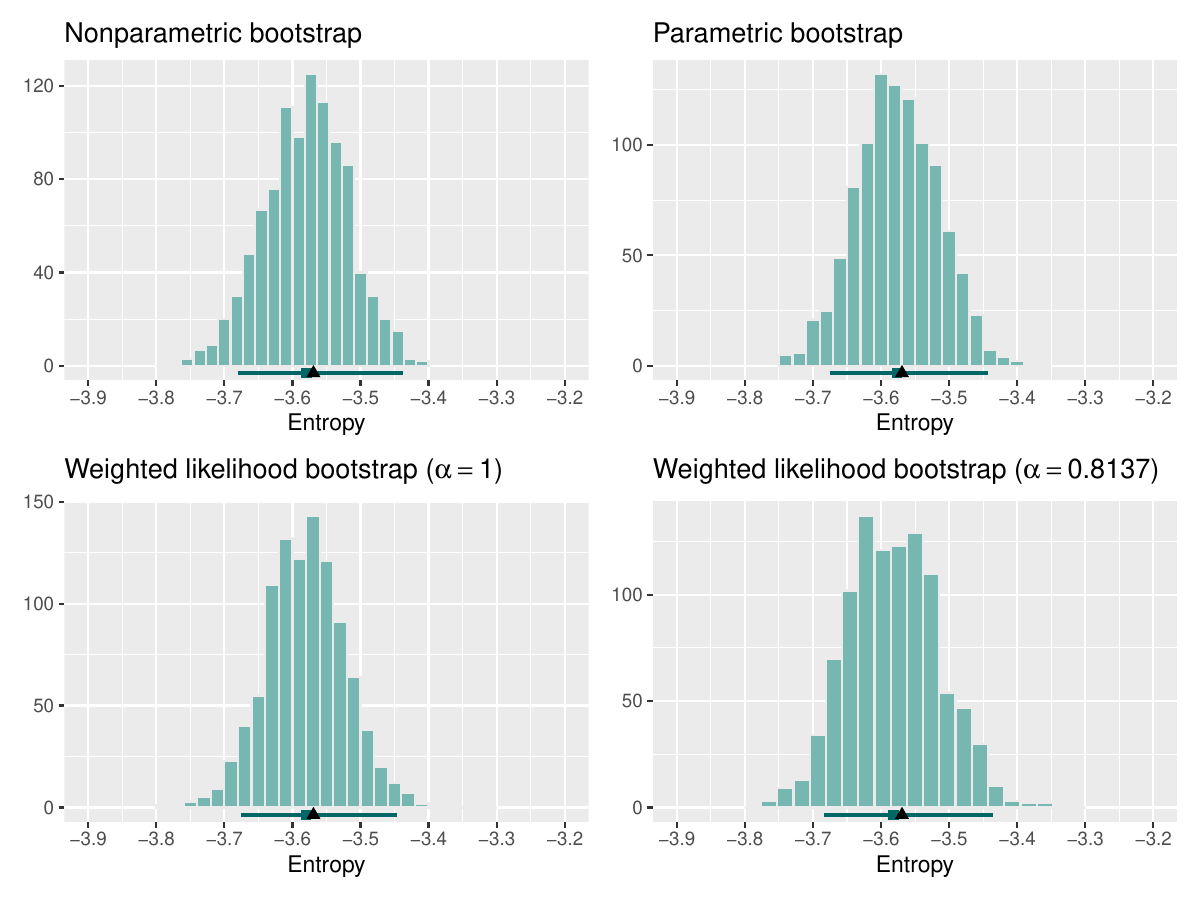}
\caption{Bootstrap distributions of entropy obtained using various bootstrap procedures for the year 2019. In each panel, filled triangles depict the GMM-based estimate of entropy, while horizontal segments represent the corresponding centered percentile intervals.}
\label{fig4:gold_prices}
\end{figure}

In Figure~\ref{fig4:gold_prices} each panel shows a histogram representing the bootstrap distribution of entropy obtained by different bootstrap methods for the year 2019. Across all panels, the bootstrap distributions appear to be relatively symmetric and spread over a similar range, between -3.8 and -3.4.
This similarity further confirms the robustness of the entropy estimates and the consistency among the various bootstrap techniques in quantifying the associated uncertainty.

Overall, the analysis reinforces the interpretation that the entropy estimates and their associated uncertainties are reliable and consistent across the different bootstrap methods employed, providing confidence in the analysis of volatility patterns in the gold market.

\clearpage

\subsection{NBA 2022-23 net rating}

Advanced basketball statistics provide deeper insights into the performance of players and teams \citep{Oliver:2004, Zuccolotto:Manisera:2020}. These metrics aim to extract meaningful patterns and relationships from data collected during basketball games. While traditional statistics such as points, rebounds, and assists provide a basic measure of a player's or team's performance, efficiency metrics, such as Effective Field Goal, True Shooting Percentage, Offensive and Defensive Rating, aim to assess the efficiency of play and the overall impact on team success.

A comprehensive measure of a team's overall game performance relative to its opponent is provided by the \emph{Net Rating} \citep{Kubatko:etal:2007}, defined as the difference between \emph{Offensive Rating} (ORtg) and \emph{Defensive Rating} (DRtg), i.e.
\begin{equation*}
\text{NetRtg} = \text{ORtg} - \text{DRtg}.
\end{equation*}
The last two measures are advanced basketball statistics used to evaluate the efficiency of teams on offense and defense, respectively.
The Offensive Rating (ORtg) is computed as:
\begin{equation*}
\text{ORtg} = \frac{ \text{Pts} }{ \text{Poss} } \times 100,
\end{equation*}
where Pts represents the total points scored by a team, and Poss refers to the total number of possessions.
The latter is calculated as:
\begin{equation*}
\text{Poss} = \text{FGA} + 0.44 \times \text{FTA} + \text{TOV} - \text{OReb},
\end{equation*}
where FGA is the number of (either 2-point and 3-point) field goal attempts, FTA is the number of free throw attempts, TOV the number of turnovers, and OReb the number of offensive rebounds.
Thus, ORtg is a measure of a team offensive efficiency, representing the number of points scored by a team per 100 possessions. A higher ORtg indicates a more efficient offensive phase, while a lower ORtg suggests less efficient offensive performance.
Defensive Rating (DRtg) is calculated similarly but refers to the opposing team. Therefore, it is a measure of a team defensive efficiency, representing the number of points allowed by the team per 100 possessions.

\begin{figure}
\centering
\includegraphics[width=0.9\textwidth]{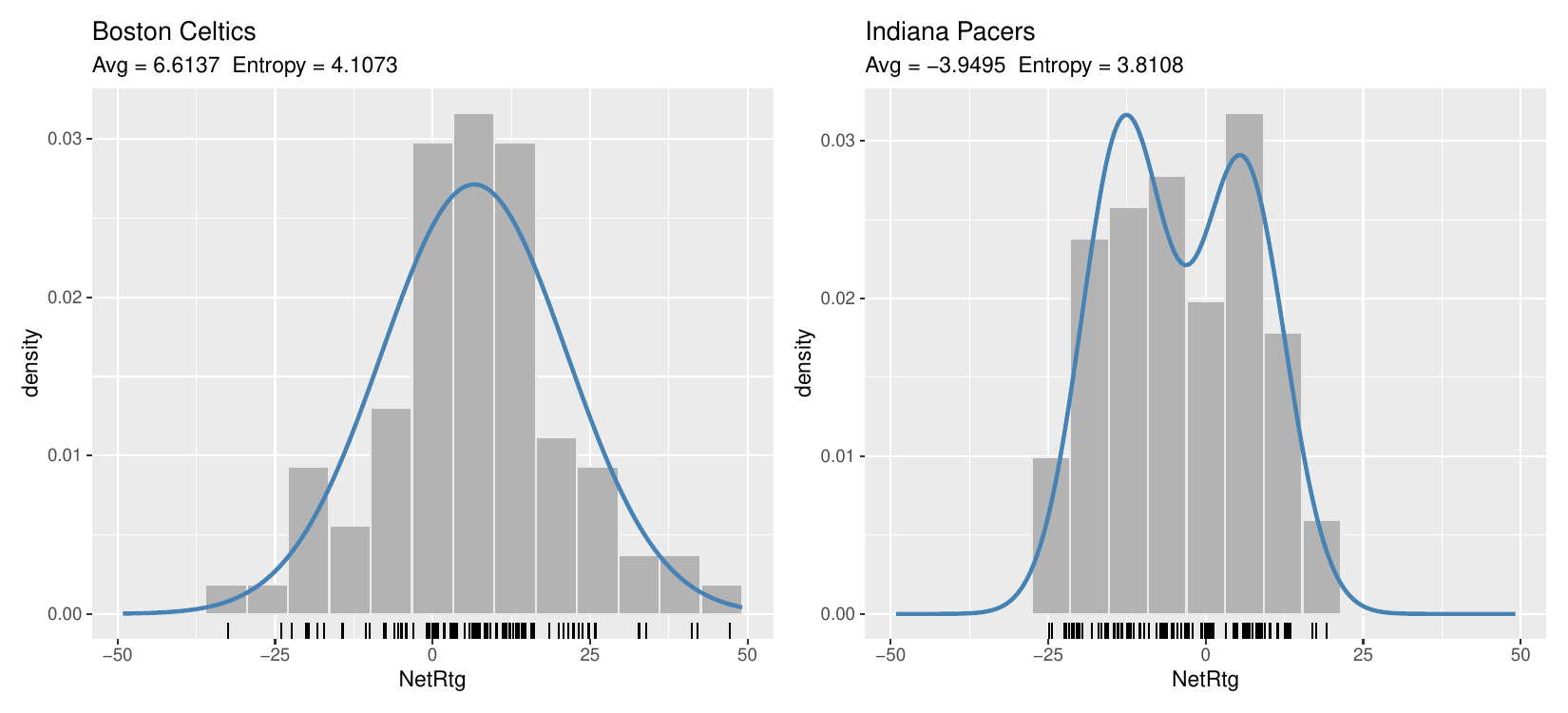}
\caption{Histogram of net rating (NetRtg) scores for two selected NBA teams with the corresponding GMM-based density estimates. Average and entropy values are also reported.}
\label{fig1:NBA2023NetRtg}
\end{figure}

We consider data collected during the 2022-23 NBA Regular Season and available through the API provided by the R package \textsf{hoopR} \citep{Rpkg:hoopR}.
The ORtg and DRtg values on each game were computed, so net rating (NetRtg) scores were obtained for each NBA team on the 82 games of the regular season. Figure~\ref{fig1:NBA2023NetRtg} shows the histograms of net rating scores for the Boston Celtics and the Indiana Pacers, with the corresponding GMM-based density estimates. The Celtics had the largest average net rating score, much larger than that of the Pacers, but also a larger entropy reflecting the higher uncertainty in the difference between team's offensive and defensive ratings.

\begin{table}[htb]
\caption{Average and entropy values of net rating for NBA teams on the 2022-23 regular season, with corresponding 95\% credible intervals obtained using the WLB approach with $\alpha=0.8137$.}
\small
\label{tab:NBA2023NetRtg}
\begin{tabular}{lcccc}
\toprule
Team & Average & 95\% CI & Entropy & 95\% CI\\
\midrule
Atlanta Hawks & -0.6568 & $[-3.9290, +2.5045]$ & 4.0469 & $[+3.8827, +4.1820]$\\
Boston Celtics & 6.6137 & $[+3.0671, +10.1934]$ & 4.1073 & $[+3.9033, +4.2879]$\\
Brooklyn Nets & -0.2344 & $[-3.8125, +3.4127]$ & 4.1757 & $[+3.9719, +4.3500]$\\
Charlotte Bobcats & -5.2925 & $[-8.4168, -1.9242]$ & 4.0099 & $[+3.8367, +4.1573]$\\
Chicago Bulls & 1.6153 & $[-1.7602, +5.2525]$ & 4.1046 & $[+3.9374, +4.2509]$\\
Cleveland Cavaliers & 4.9818 & $[+1.7502, +8.7335]$ & 4.0994 & $[+3.9419, +4.2187]$\\
Dallas Mavericks & -0.3520 & $[-3.2899, +2.8415]$ & 3.9602 & $[+3.7321, +4.1800]$\\
Denver Nuggets & 3.4118 & $[-0.1471, +6.6683]$ & 4.1011 & $[+3.9228, +4.2544]$\\
Detroit Pistons & -8.4807 & $[-11.4003, -5.2130]$ & 3.9897 & $[+3.8319, +4.1230]$\\
Golden State Warriors & 0.7551 & $[-3.4145, +4.6851]$ & 4.2258 & $[+4.0118, +4.4048]$\\
Houston Rockets & -6.4060 & $[-9.4581, -3.3076]$ & 4.0143 & $[+3.8274, +4.1719]$\\
Indiana Pacers & -3.9495 & $[-7.1604, -0.9001]$ & 3.8108 & $[+3.6811, +3.8880]$\\
LA Clippers & 0.7368 & $[-2.4117, +4.0755]$ & 4.0442 & $[+3.8849, +4.1700]$\\
Los Angeles Lakers & 0.9280 & $[-2.0730, +3.9052]$ & 3.9679 & $[+3.7996, +4.1262]$\\
Memphis Grizzlies & 5.0635 & $[+1.5102, +8.6959]$ & 4.1319 & $[+3.9694, +4.2848]$\\
Miami Heat & -0.3530 & $[-3.2717, +2.5170]$ & 3.8992 & $[+3.6811, +4.0630]$\\
Milwaukee Bucks & 5.2019 & $[+0.0575, +8.7758]$ & 4.1030 & $[+3.9004, +4.2525]$\\
Minnesota Timberwolves & -0.5680 & $[-3.6957, +2.4405]$ & 3.9862 & $[+3.8350, +4.1170]$\\
New Orleans Hornets & 2.0084 & $[-1.4134, +5.7065]$ & 4.1758 & $[+4.0204, +4.3205]$\\
New York Knicks & 3.8174 & $[+0.1951, +7.3190]$ & 3.9624 & $[+3.7467, +4.1145]$\\
Oklahoma City Thunder & 0.2574 & $[-2.9938, +3.5134]$ & 4.0478 & $[+3.8801, +4.1798]$\\
Orlando Magic & -2.0689 & $[-5.0916, +1.1678]$ & 3.9747 & $[+3.8219, +4.1131]$\\
Philadelphia 76ers & 3.6097 & $[+0.2767, +7.2501]$ & 4.0897 & $[+3.9346, +4.2499]$\\
Phoenix Suns & 1.8343 & $[-1.8363, +5.6421]$ & 4.2162 & $[+4.0527, +4.3465]$\\
Portland Trail Blazers & -4.7156 & $[-8.8471, -1.0848]$ & 4.1898 & $[+4.0327, +4.3369]$\\
Sacramento Kings & 3.1652 & $[-0.4250, +6.4799]$ & 4.0896 & $[+3.9118, +4.2613]$\\
San Antonio Spurs & -10.4663 & $[-14.1456, -6.9153]$ & 4.1707 & $[+4.0208, +4.2829]$\\
Toronto Raptors & 1.3072 & $[-1.7371, +4.7026]$ & 3.9851 & $[+3.8263, +4.1479]$\\
Utah Jazz & -1.1229 & $[-4.1182, +1.8446]$ & 3.9444 & $[+3.7549, +4.0888]$\\
Washington Wizards & -0.6406 & $[-3.9708, +2.6205]$ & 4.0346 & $[+3.8766, +4.1670]$\\
\bottomrule
\end{tabular}
\end{table}

\begin{figure}[htb]
\centering
\includegraphics[width=0.9\textwidth]{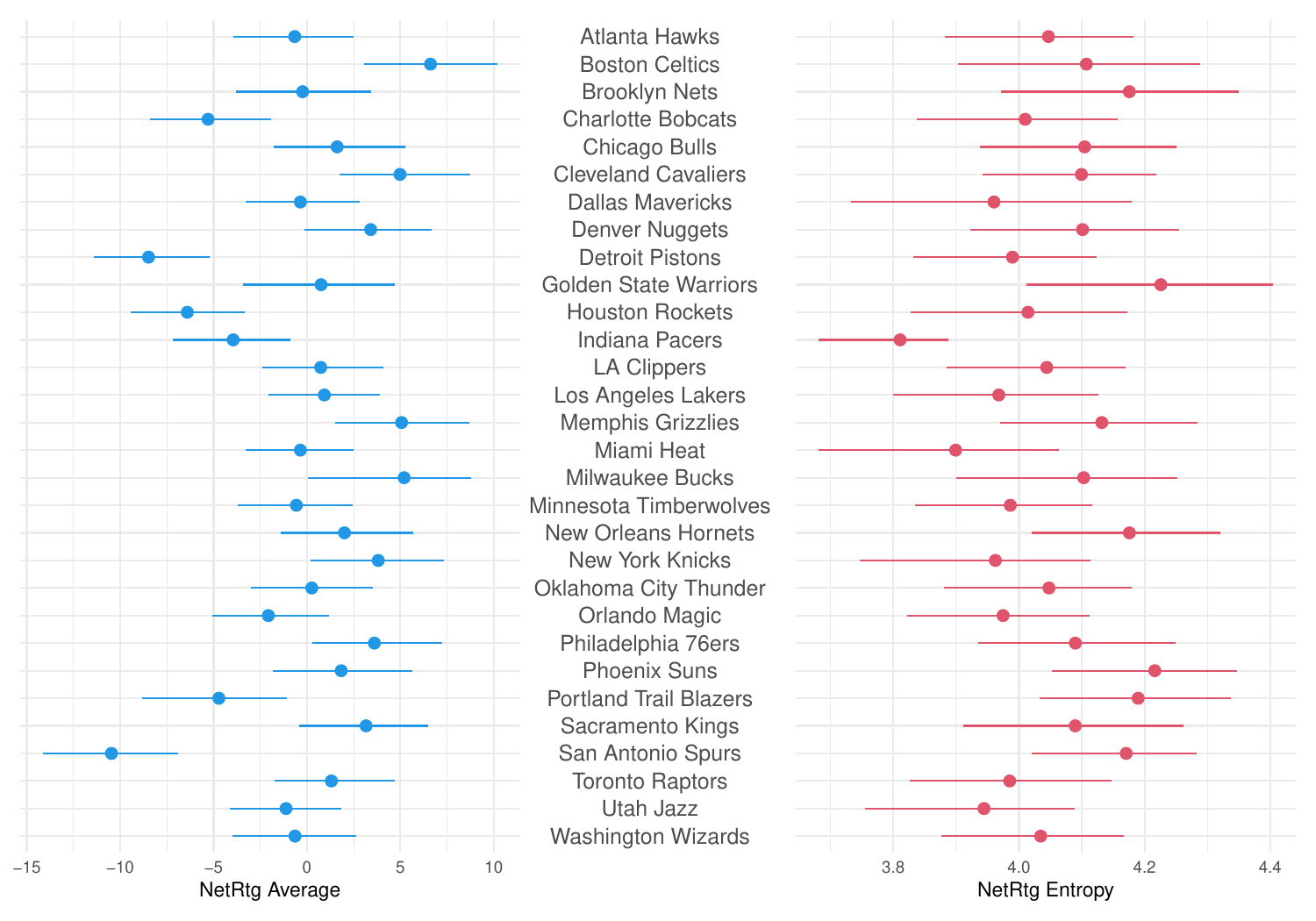}
\caption{Dotcharts of net rating (NetRtg) average and entropy for NBA teams on the 2022-23 regular season, with corresponding 95\% credible intervals obtained using the WLB approach with $\alpha = 0.8137$.}
\label{fig2:NBA2023NetRtg}
\end{figure}

Table~\ref{tab:NBA2023NetRtg} reports the average net rating for all the NBA teams, with the corresponding 95\% confidence intervals obtained using the WLB approach as described in \citet{Scrucca:etal:2016} but with $\alpha=0.8137$ as described in Section~\ref{sec:wlbwts}.
The table also includes the estimated entropy and the 95\% confidence intervals computed according to the WLB with $\alpha=0.8137$.
The same information is also shown graphically in Figure~\ref{fig2:NBA2023NetRtg}.

The NBA teams with the lowest net rating averages are those worst ranked in the final 2022-23 regular season standings. In contrast, the Boston Celtics (BOS) and Milwaukee Bucks (MIL) show the highest averages and were the top two teams in the final regular season rankings.
The San Antonio Spurs (SA) have both the lowest average and one of the highest entropy, indicating great fluctuation in their NetRtg statistics.
The team with the highest entropy is the Golden State Warriors (GS), which showed large fluctuations, with net ratings going from -50 to +50. This is twice the values observed for the Indiana Pacers (IND), which on the contrary have the smallest entropy.

Overall, for a comprehensive assessment of a team's performance, it is important to consider both the average net rating and its entropy. The net rating average describes the overall quality and performance level of an NBA team, with higher values indicating stronger teams. In contrast, net rating entropy measures the consistency of a team's performance across the season, with higher entropy values indicating greater fluctuation from game to game.
Ideally, successful teams should aim for both a high positive net rating, reflecting strong overall performance, and low entropy, indicating consistent play. 

\clearpage
\section{Conclusion}
\label{sec:conc}

In this paper, we presented a novel framework for assessing the uncertainty associated with mixture-based entropy estimation.
Our proposal exploits the underlying mixture structure by assigning random weights to observations in a weighted likelihood bootstrap (WLB) procedure, leading to more accurate uncertainty quantification.
Through extensive simulation studies, we compared the performance of different resampling strategies, including nonparametric bootstrap, parametric bootstrap, and WLB with varying weight generation schemes. The results demonstrated that the WLB approach with weights generated from a Dirichlet distribution with parameter $\alpha = 0.8137$ consistently provided empirical coverage levels closest to the nominal level across a wide range of scenarios. Additionally, the use of centered percentile intervals emerged as the preferred choice to ensure reliable empirical coverage.
We illustrated the practical utility of our proposed method by analyzing two real-world datasets: daily log-returns of gold prices at COMEX from 2014 to 2022, and the Net Rating scores for NBA teams during the 2022/23 regular season.
These applications have shown the effectiveness of our approach in quantifying uncertainty in entropy estimation and providing insights into the volatility patterns and performance consistency of these datasets.

Overall, our work contributes to the field of entropy estimation by proposing a novel methodology that addresses the crucial need for reliable uncertainty assessment in mixture-based entropy estimation. The proposed WLB approach with optimized weight generation and the use of centered percentile intervals offer a robust and accurate framework for quantifying uncertainty, enabling more informed decision-making and deeper insights into the underlying data distributions.

Future research directions could focus on extending the WLB approach for uncertainty assessment of entropy estimation in more complex data structures beyond mixture models, such as time series, spatial processes, and data with hierarchical or multilevel dependencies.


\section*{Acknowledgements}
I would like to express my gratitude to the Istituto Nazionale di Fisica Nucleare (INFN) for providing the scientific computing infrastructure used for running the simulations presented in this work.

\clearpage
\bibliography{paper_arxiv}

\clearpage
\appendix

\section*{Appendix}

\section{Simulation results}
\label{ap:simres}

\begin{table}[h]
\caption{Simulation results for data generated from a Gaussian distribution with $\mu = 0$ and $\sigma = 1$. For each bootstrap method and sample size, the table reports averaged values of the entropy estimates, bootstrap estimates of bias, standard error, and percentile interval limits, along with their empirical coverage.}
\label{tab:Gaussian_mu=0_sigma=1}
\small
\centering
\addtolength{\tabcolsep}{-0.6ex}
\begin{tabular}{lccccccccc}
\toprule
 & & & & \multicolumn{3}{c}{95\% percentile interval} & \multicolumn{3}{c}{95\% centered perc. interval} \\
\cmidrule(l{3pt}r{3pt}){5-7} \cmidrule(l{3pt}r{3pt}){8-10}
Sample size & \textbf{Estimate} & \textbf{Bias} & \textbf{SE} & lower & upper & \textbf{Coverage} & lower & upper & \textbf{Coverage}\\
\midrule
\multicolumn{10}{l}{Nonparametric bootstrap}\\[1ex]
 100 & 1.4096 & 0.0104 & 0.0706 & 1.2556 & 1.5305 & 0.925 & 1.2886 & 1.5632 & 0.945\\
 200 & 1.4141 & 0.0050 & 0.0498 & 1.3092 & 1.5035 & 0.942 & 1.3248 & 1.5191 & 0.945\\
 500 & 1.4163 & 0.0020 & 0.0316 & 1.3517 & 1.4750 & 0.936 & 1.3575 & 1.4810 & 0.944\\
1000 & 1.4170 & 0.0011 & 0.0224 & 1.3719 & 1.4592 & 0.955 & 1.3749 & 1.4621 & 0.957\\
\midrule
\multicolumn{10}{l}{Parametric bootstrap}\\[1ex]
100 & 1.4096 & 0.0105 & 0.0715 & 1.2548 & 1.5338 & 0.939 & 1.2853 & 1.5637 & 0.951\\
200 & 1.4141 & 0.0050 & 0.0502 & 1.3088 & 1.5047 & 0.943 & 1.3236 & 1.5195 & 0.947\\
500 & 1.4163 & 0.0020 & 0.0316 & 1.3516 & 1.4751 & 0.944 & 1.3574 & 1.4809 & 0.947\\
1000 & 1.4170 & 0.0010 & 0.0224 & 1.3718 & 1.4593 & 0.957 & 1.3748 & 1.4621 & 0.962\\
\midrule
\multicolumn{10}{l}{Weighted likelihood bootstrap ($\alpha = 1$)}\\[1ex]
 100 & 1.4096 & 0.0101 & 0.0684 & 1.2675 & 1.5345 & 0.930 & 1.2851 & 1.5516 & 0.948\\
 200 & 1.4141 & 0.0049 & 0.0490 & 1.3148 & 1.5062 & 0.939 & 1.3219 & 1.5138 & 0.943\\
 500 & 1.4163 & 0.0020 & 0.0314 & 1.3537 & 1.4762 & 0.946 & 1.3564 & 1.4788 & 0.939\\
1000 & 1.4170 & 0.0010 & 0.0223 & 1.3729 & 1.4599 & 0.960 & 1.3742 & 1.4612 & 0.957\\
\midrule
\multicolumn{10}{l}{Weighted likelihood bootstrap ($\alpha = 4$)}\\[1ex]
 100 & 1.4096 & 0.0025 & 0.0346 & 1.3401 & 1.4753 & 0.654 & 1.3441 & 1.4792 & 0.650\\
 200 & 1.4141 & 0.0013 & 0.0247 & 1.3650 & 1.4615 & 0.694 & 1.3668 & 1.4633 & 0.689\\
 500 & 1.4163 & 0.0006 & 0.0157 & 1.3854 & 1.4467 & 0.664 & 1.3859 & 1.4474 & 0.667\\
1000 & 1.4170 & 0.0003 & 0.0111 & 1.3951 & 1.4387 & 0.681 & 1.3955 & 1.4389 & 0.684\\
\midrule
\multicolumn{10}{l}{Weighted likelihood bootstrap ($\alpha = 0.8137$)}\\[1ex]
 100 & 1.4096 & 0.0121 & 0.0754 & 1.2517 & 1.5460 & 0.954 & 1.2729 & 1.5670 & 0.962\\
 200 & 1.4141 & 0.0061 & 0.0541 & 1.3038 & 1.5152 & 0.963 & 1.3129 & 1.5244 & 0.966\\
 500 & 1.4163 & 0.0024 & 0.0347 & 1.3471 & 1.4825 & 0.958 & 1.3503 & 1.4855 & 0.965\\
1000 & 1.4170 & 0.0013 & 0.0246 & 1.3684 & 1.4644 & 0.973 & 1.3698 & 1.4658 & 0.971\\
\bottomrule
\end{tabular}
\end{table}

\begin{figure}
\centering
\includegraphics[width=\textwidth]{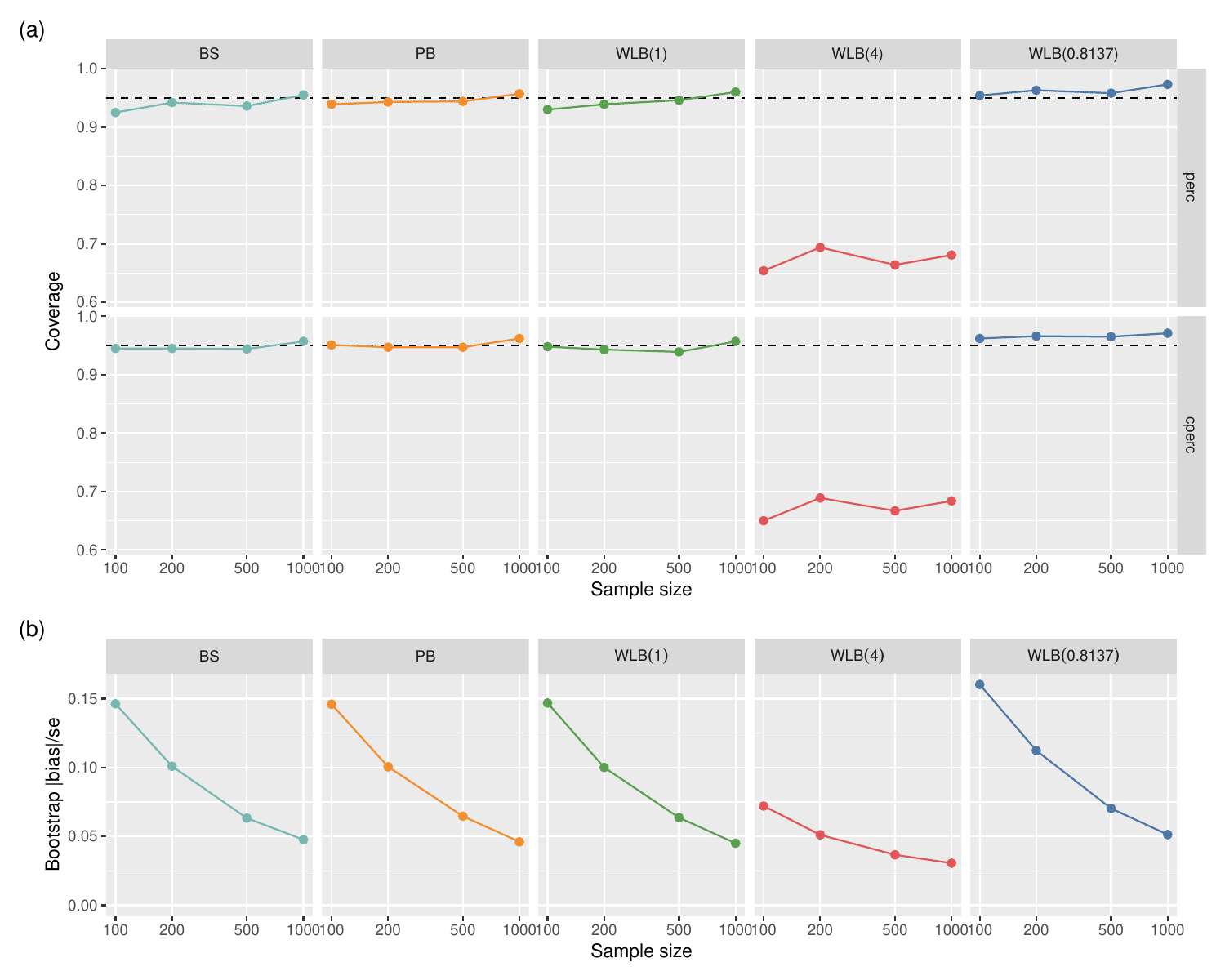}
\caption{Simulation results for data generated from a Gaussian distribution with $\mu = 0$ and $\sigma = 1$.
Panel (a) shows the empirical coverage for 1000 replications as function of the sample size for different bootstrap methods (BS = nonparametric; PB = parametric; WLB = weighted likelihood bootstrap with weights generated from a Dirichlet with different values of $\alpha$) and two type of bootstrap intervals (perc = percentile; cperc = centered percentile).
Panel (b) show the ratio of bias (in absolute value) over the standard error for each bootstrap procedure as function of sample size.}
\label{fig:Gaussian_mu=0_sigma=1}
\end{figure}

\begin{table}
\caption{Simulation results for data generated from a Mixed Gaussian distribution with $\mu = 2$ and $\sigma = 1$. For each bootstrap method and sample size, the table reports averaged values of the entropy estimates, bootstrap estimates of bias, standard error, and percentile interval limits, along with their empirical coverage.}
\label{tab:MixedGaussian_mu=2_sigma=1}
\small
\centering
\addtolength{\tabcolsep}{-0.6ex}
\begin{tabular}{lccccccccc}
\toprule
 & & & & \multicolumn{3}{c}{95\% percentile interval} & \multicolumn{3}{c}{95\% centered perc. interval} \\
\cmidrule(l{3pt}r{3pt}){5-7} \cmidrule(l{3pt}r{3pt}){8-10}
Sample size & \textbf{Estimate} & \textbf{Bias} & \textbf{SE} & lower & upper & \textbf{Coverage} & lower & upper & \textbf{Coverage}\\
\midrule
\multicolumn{10}{l}{Nonparametric bootstrap}\\[1ex]
 100 & 2.0273 & 0.0205 & 0.0601 & 1.8821 & 2.1158 & 0.861 & 1.9388 & 2.1728 & 0.936\\
 200 & 2.0436 & 0.0100 & 0.0413 & 1.9495 & 2.1105 & 0.901 & 1.9765 & 2.1379 & 0.924\\
 500 & 2.0474 & 0.0040 & 0.0259 & 1.9915 & 2.0926 & 0.941 & 2.0021 & 2.1034 & 0.961\\
1000 & 2.0511 & 0.0020 & 0.0183 & 2.0128 & 2.0841 & 0.955 & 2.0181 & 2.0895 & 0.962\\
\midrule
\multicolumn{10}{l}{Parametric bootstrap}\\[1ex]
 100 & 2.0273 & 0.0206 & 0.0613 & 1.8805 & 2.1198 & 0.880 & 1.9350 & 2.1736 & 0.943\\
 200 & 2.0436 & 0.0101 & 0.0420 & 1.9484 & 2.1123 & 0.911 & 1.9750 & 2.1386 & 0.934\\
 500 & 2.0474 & 0.0042 & 0.0261 & 1.9911 & 2.0929 & 0.949 & 2.0018 & 2.1036 & 0.961\\
1000 & 2.0511 & 0.0024 & 0.0183 & 2.0128 & 2.0841 & 0.953 & 2.0180 & 2.0895 & 0.955\\
\midrule
\multicolumn{10}{l}{Weighted likelihood bootstrap ($\alpha = 1$)}\\[1ex]
 100 & 2.0273 & 0.0199 & 0.0577 & 1.8920 & 2.1175 & 0.865 & 1.9368 & 2.1625 & 0.934\\
 200 & 2.0436 & 0.0099 & 0.0405 & 1.9542 & 2.1124 & 0.896 & 1.9749 & 2.1332 & 0.923\\
 500 & 2.0474 & 0.0040 & 0.0257 & 1.9933 & 2.0935 & 0.942 & 2.0010 & 2.1014 & 0.957\\
1000 & 2.0511 & 0.0020 & 0.0182 & 2.0137 & 2.0847 & 0.954 & 2.0175 & 2.0885 & 0.957\\
\midrule
\multicolumn{10}{l}{Weighted likelihood bootstrap ($\alpha = 4$)}\\[1ex]
 100 & 2.0273 & 0.0050 & 0.0290 & 1.9652 & 2.0783 & 0.623 & 1.9764 & 2.0894 & 0.666\\
 200 & 2.0436 & 0.0025 & 0.0203 & 2.0014 & 2.0806 & 0.627 & 2.0064 & 2.0859 & 0.637\\
 500 & 2.0474 & 0.0010 & 0.0129 & 2.0213 & 2.0716 & 0.697 & 2.0232 & 2.0734 & 0.692\\
1000 & 2.0511 & 0.0005 & 0.0091 & 2.0329 & 2.0684 & 0.698 & 2.0337 & 2.0694 & 0.696\\
\midrule
\multicolumn{10}{l}{Weighted likelihood bootstrap ($\alpha = 0.8137$)}\\[1ex]
 100 & 2.0273 & 0.0243 & 0.0639 & 1.8751 & 2.1247 & 0.885 & 1.9305 & 2.1793 & 0.947\\
 200 & 2.0436 & 0.0122 & 0.0448 & 1.9430 & 2.1182 & 0.924 & 1.9690 & 2.1440 & 0.952\\
 500 & 2.0474 & 0.0049 & 0.0284 & 1.9870 & 2.0979 & 0.959 & 1.9967 & 2.1077 & 0.976\\
1000 & 2.0511 & 0.0024 & 0.0201 & 2.0095 & 2.0881 & 0.973 & 2.0141 & 2.0927 & 0.968\\
\bottomrule
\end{tabular}
\end{table}

\begin{figure}
\centering
\includegraphics[width=\textwidth]{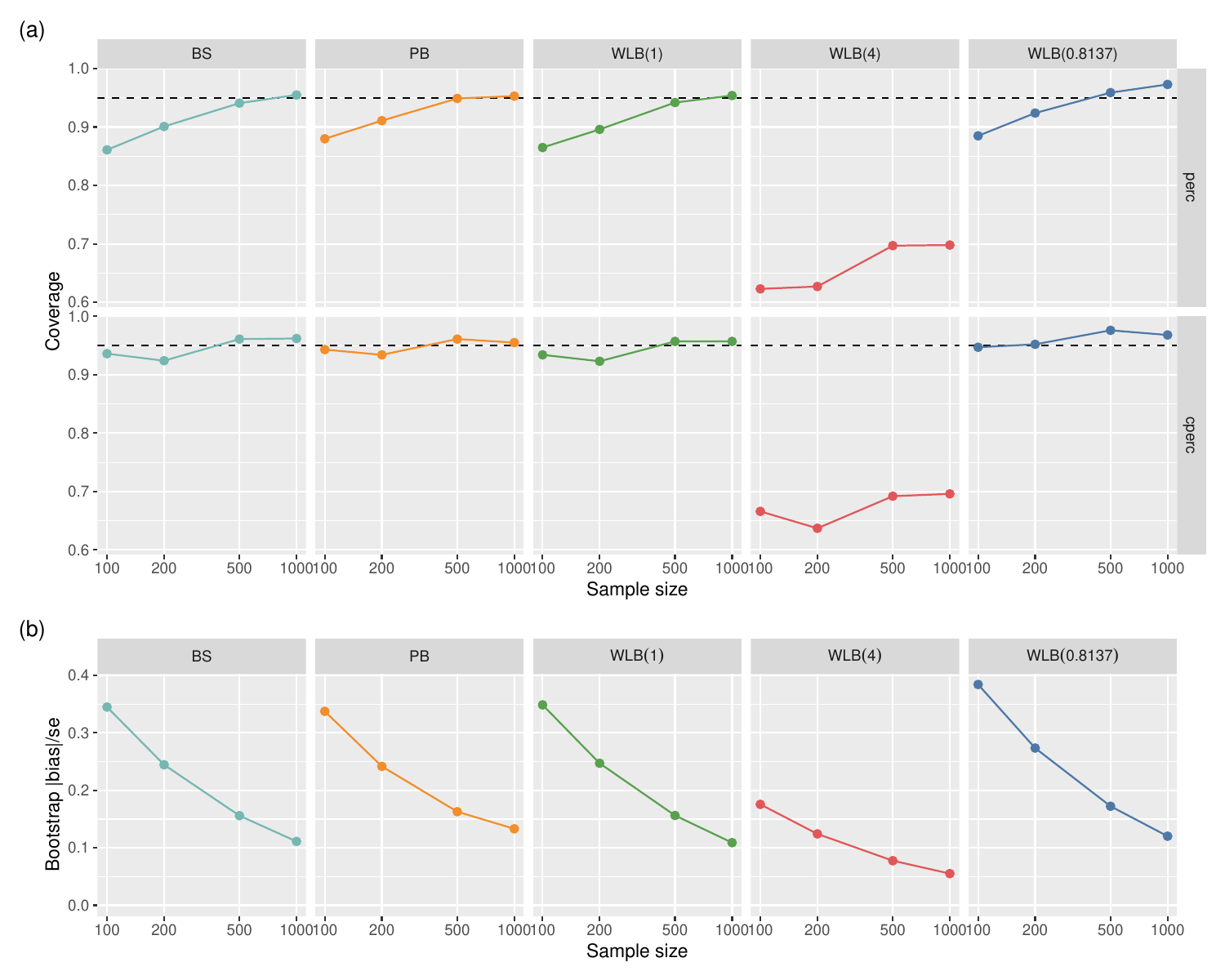}
\caption{Simulation results for data generated from a Mixed Gaussian distribution with $\mu = 2$ and $\sigma = 1$.
Panel (a) shows the empirical coverage for 1000 replications as function of the sample size for different bootstrap methods (BS = nonparametric; PB = parametric; WLB = weighted likelihood bootstrap with weights generated from a Dirichlet with different values of $\alpha$) and two type of bootstrap intervals (perc = percentile; cperc = centered percentile).
Panel (b) show the ratio of bias (in absolute value) over the standard error for each bootstrap procedure as function of sample size.}
\label{fig:MixedGaussian_mu=2_sigma=1}
\end{figure}

\begin{table}
\caption{Simulation results for data generated from a Laplace distribution with $\mu = 0$ and $\beta = 2$. For each bootstrap method and sample size, the table reports averaged values of the entropy estimates, bootstrap estimates of bias, standard error, and percentile interval limits, along with their empirical coverage.}
\label{tab:Laplace_mu=0_beta=2}
\small
\centering
\addtolength{\tabcolsep}{-0.6ex}
\begin{tabular}{lccccccccc}
\toprule
 & & & & \multicolumn{3}{c}{95\% percentile interval} & \multicolumn{3}{c}{95\% centered perc. interval} \\
\cmidrule(l{3pt}r{3pt}){5-7} \cmidrule(l{3pt}r{3pt}){8-10}
Sample size & \textbf{Estimate} & \textbf{Bias} & \textbf{SE} & lower & upper & \textbf{Coverage} & lower & upper & \textbf{Coverage}\\
\midrule
\multicolumn{10}{l}{Nonparametric bootstrap}\\[1ex]
 100 & 2.3742 & 0.0259 & 0.1020 & 2.1380 & 2.5351 & 0.904 & 2.2134 & 2.6105 & 0.919\\
 200 & 2.3767 & 0.0154 & 0.0709 & 2.2193 & 2.4955 & 0.926 & 2.2581 & 2.5345 & 0.945\\
 500 & 2.3858 & 0.0066 & 0.0447 & 2.2909 & 2.4652 & 0.945 & 2.3066 & 2.4809 & 0.948\\
1000 & 2.3870 & 0.0035 & 0.0317 & 2.3212 & 2.4449 & 0.947 & 2.3292 & 2.4530 & 0.941\\
\midrule
\multicolumn{10}{l}{Parametric bootstrap}\\[1ex]
 100 & 2.3742 & 0.0201 & 0.0899 & 2.1725 & 2.5234 & 0.870 & 2.2249 & 2.5758 & 0.883\\
 200 & 2.3767 & 0.0118 & 0.0676 & 2.2302 & 2.4938 & 0.918 & 2.2592 & 2.5231 & 0.930\\
 500 & 2.3858 & 0.0042 & 0.0436 & 2.2955 & 2.4655 & 0.947 & 2.3061 & 2.4762 & 0.948\\
1000 & 2.3870 & 0.0025 & 0.0310 & 2.3243 & 2.4450 & 0.941 & 2.3291 & 2.4500 & 0.942\\
\midrule
\multicolumn{10}{l}{Weighted likelihood bootstrap ($\alpha = 1$)}\\[1ex]
 100 & 2.3742 & 0.0226 & 0.0958 & 2.1659 & 2.5384 & 0.899 & 2.2101 & 2.5829 & 0.917\\
 200 & 2.3767 & 0.0144 & 0.0689 & 2.2287 & 2.4976 & 0.922 & 2.2560 & 2.5246 & 0.938\\
 500 & 2.3858 & 0.0065 & 0.0441 & 2.2940 & 2.4661 & 0.944 & 2.3055 & 2.4778 & 0.948\\
1000 & 2.3870 & 0.0034 & 0.0315 & 2.3226 & 2.4454 & 0.950 & 2.3285 & 2.4515 & 0.946\\
\midrule
\multicolumn{10}{l}{Weighted likelihood bootstrap ($\alpha = 4$)}\\[1ex]
 100 & 2.3742 & 0.0057 & 0.0481 & 2.2750 & 2.4627 & 0.616 & 2.2853 & 2.4731 & 0.639\\
 200 & 2.3767 & 0.0038 & 0.0346 & 2.3057 & 2.4408 & 0.629 & 2.3125 & 2.4478 & 0.635\\
 500 & 2.3858 & 0.0017 & 0.0222 & 2.3412 & 2.4276 & 0.675 & 2.3439 & 2.4305 & 0.682\\
1000 & 2.3870 & 0.0009 & 0.0158 & 2.3554 & 2.4170 & 0.677 & 2.3569 & 2.4186 & 0.679\\
\midrule
\multicolumn{10}{l}{Weighted likelihood bootstrap ($\alpha = 0.8137$)}\\[1ex]
 100 & 2.3742 & 0.0276 & 0.1060 & 2.1403 & 2.5520 & 0.919 & 2.1956 & 2.6082 & 0.935\\
 200 & 2.3767 & 0.0175 & 0.0761 & 2.2113 & 2.5083 & 0.951 & 2.2451 & 2.5418 & 0.961\\
 500 & 2.3858 & 0.0079 & 0.0487 & 2.2838 & 2.4737 & 0.964 & 2.2977 & 2.4878 & 0.971\\
1000 & 2.3870 & 0.0042 & 0.0348 & 2.3153 & 2.4512 & 0.968 & 2.3226 & 2.4584 & 0.967\\
\bottomrule
\end{tabular}
\end{table}

\begin{figure}
\centering
\includegraphics[width=\textwidth]{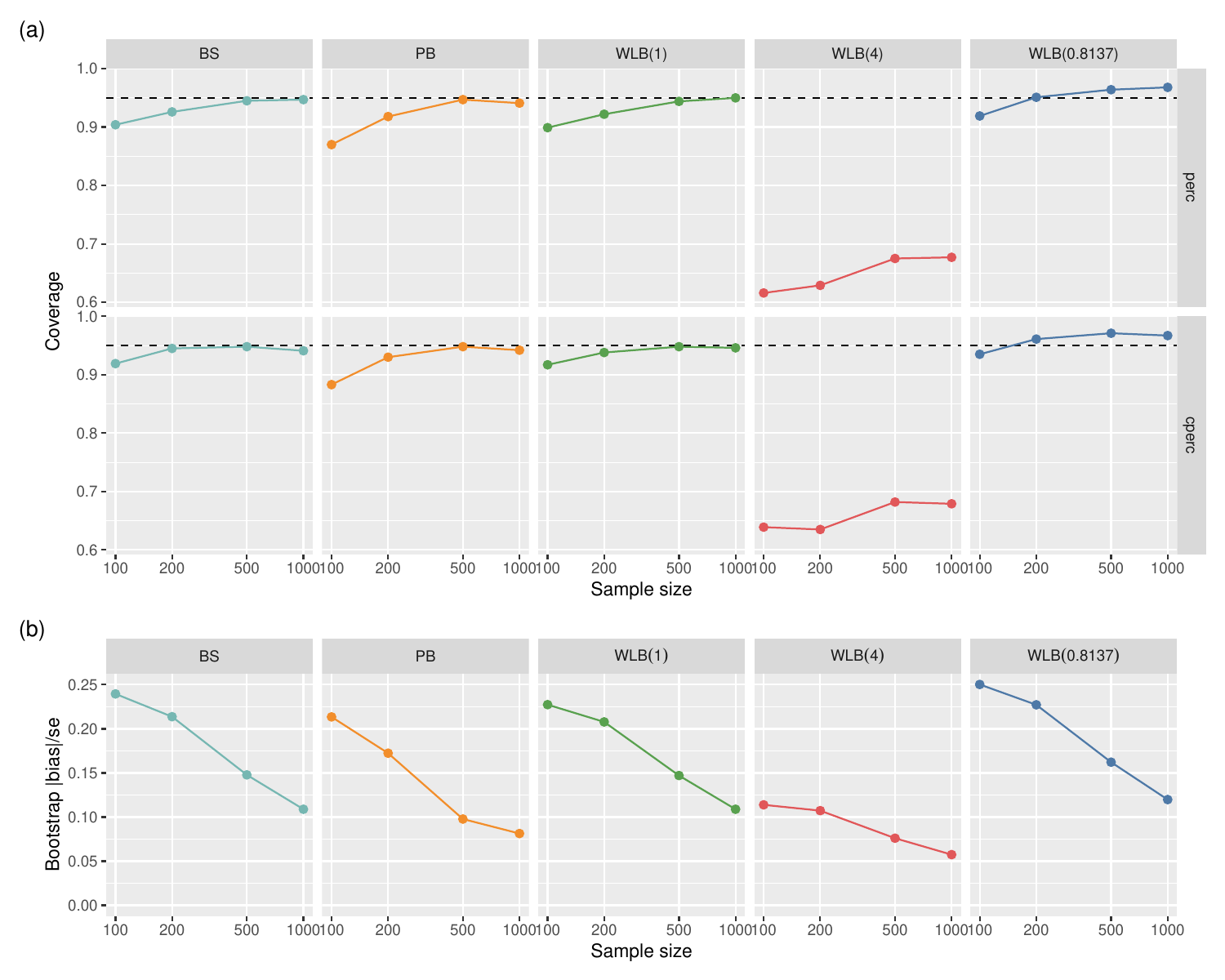}
\caption{Simulation results for data generated from a Laplace distribution with $\mu = 0$ and $\beta = 2$.
Panel (a) shows the empirical coverage for 1000 replications as function of the sample size for different bootstrap methods (BS = nonparametric; PB = parametric; WLB = weighted likelihood bootstrap with weights generated from a Dirichlet with different values of $\alpha$) and two type of bootstrap intervals (perc = percentile; cperc = centered percentile).
Panel (b) show the ratio of bias (in absolute value) over the standard error for each bootstrap procedure as function of sample size.}
\label{fig:Laplace_mu=0_beta=2}
\end{figure}

\begin{table}
\caption{Simulation results for data generated from a bivariate Gaussian distribution with $\mu = [0,0]\T$ and ${\Sigmab = \begin{bsmatrix} 1.0 & 0.8 \\ 0.8 & 2.0 \end{bsmatrix}}$. For each bootstrap method and sample size, the table reports averaged values of the entropy estimates, bootstrap estimates of bias, standard error, and percentile interval limits, along with their empirical coverage.}
\label{tab:BivGaussian}
\small
\centering
\addtolength{\tabcolsep}{-0.6ex}
\begin{tabular}{lccccccccc}
\toprule
 & & & & \multicolumn{3}{c}{95\% percentile interval} & \multicolumn{3}{c}{95\% centered perc. interval} \\
\cmidrule(l{3pt}r{3pt}){5-7} \cmidrule(l{3pt}r{3pt}){8-10}
Sample size & \textbf{Estimate} & \textbf{Bias} & \textbf{SE} & lower & upper & \textbf{Coverage} & lower & upper & \textbf{Coverage}\\
\midrule
\multicolumn{10}{l}{Nonparametric bootstrap}\\[1ex]
 100 & 2.9634 & 0.0258 & 0.0992 & 2.7375 & 3.1241 & 0.896 & 2.8029 & 3.1894 & 0.940\\
 200 & 2.9779 & 0.0126 & 0.0702 & 2.8254 & 3.0994 & 0.915 & 2.8565 & 3.1303 & 0.923\\
 500 & 2.9829 & 0.0050 & 0.0447 & 2.8898 & 3.0641 & 0.940 & 2.9019 & 3.0761 & 0.958\\
1000 & 2.9896 & 0.0025 & 0.0315 & 2.9252 & 3.0481 & 0.928 & 2.9312 & 3.0541 & 0.936\\
\midrule
\multicolumn{10}{l}{Parametric bootstrap}\\[1ex]
 100 & 2.9634 & 0.0260 & 0.1014 & 2.7354 & 3.1305 & 0.910 & 2.7964 & 3.1918 & 0.943\\
 200 & 2.9779 & 0.0125 & 0.0712 & 2.8242 & 3.1021 & 0.923 & 2.8541 & 3.1315 & 0.933\\
 500 & 2.9829 & 0.0051 & 0.0448 & 2.8897 & 3.0644 & 0.943 & 2.9014 & 3.0765 & 0.950\\
1000 & 2.9896 & 0.0025 & 0.0316 & 2.9249 & 3.0485 & 0.934 & 2.9309 & 3.0545 & 0.940\\
\midrule
\multicolumn{10}{l}{Weighted likelihood bootstrap ($\alpha = 1$)}\\[1ex]
 100 & 2.9634 & 0.0250 & 0.0956 & 2.7517 & 3.1248 & 0.896 & 2.8022 & 3.1750 & 0.927\\
 200 & 2.9779 & 0.0122 & 0.0687 & 2.8325 & 3.1011 & 0.915 & 2.8547 & 3.1233 & 0.923\\
 500 & 2.9829 & 0.0049 & 0.0443 & 2.8924 & 3.0652 & 0.942 & 2.9008 & 3.0738 & 0.951\\
1000 & 2.9896 & 0.0025 & 0.0314 & 2.9263 & 3.0487 & 0.927 & 2.9304 & 3.0531 & 0.937\\
\midrule
\multicolumn{10}{l}{Weighted likelihood bootstrap ($\alpha = 4$)}\\[1ex]
 100 & 2.9634 & 0.0062 & 0.0484 & 2.8630 & 3.0519 & 0.595 & 2.8752 & 3.0637 & 0.616\\
 200 & 2.9779 & 0.0031 & 0.0347 & 2.9075 & 3.0428 & 0.635 & 2.9127 & 3.0485 & 0.640\\
 500 & 2.9829 & 0.0013 & 0.0223 & 2.9384 & 3.0253 & 0.664 & 2.9405 & 3.0274 & 0.660\\
1000 & 2.9896 & 0.0007 & 0.0157 & 2.9585 & 3.0198 & 0.671 & 2.9595 & 3.0208 & 0.667\\
\midrule
\multicolumn{10}{l}{Weighted likelihood bootstrap ($\alpha = 0.8137$)}\\[1ex]
 100 & 2.9634 & 0.0302 & 0.1057 & 2.7261 & 3.1381 & 0.918 & 2.7884 & 3.2002 & 0.953\\
 200 & 2.9779 & 0.0149 & 0.0758 & 2.8162 & 3.1124 & 0.937 & 2.8432 & 3.1401 & 0.950\\
 500 & 2.9829 & 0.0061 & 0.0489 & 2.8821 & 3.0732 & 0.961 & 2.8927 & 3.0839 & 0.970\\
1000 & 2.9896 & 0.0030 & 0.0347 & 2.9194 & 3.0548 & 0.953 & 2.9244 & 3.0599 & 0.950\\
\bottomrule
\end{tabular}
\end{table}

\begin{figure}
\centering
\includegraphics[width=\textwidth]{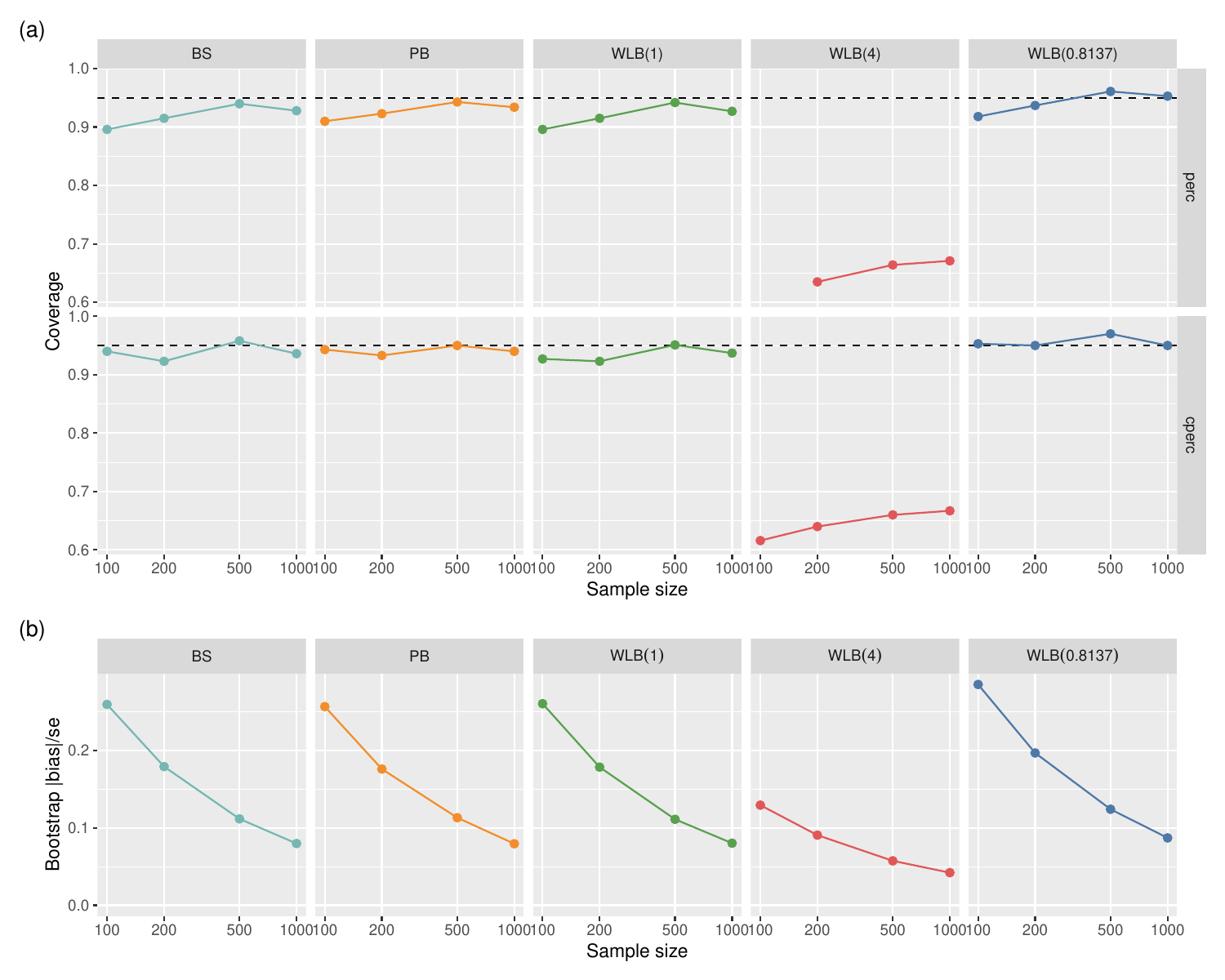}
\caption{Simulation results for data generated from a bivariate Gaussian distribution with $\mu = [0,0]\T$ and ${\Sigmab = \begin{bsmatrix} 1.0 & 0.8 \\ 0.8 & 2.0 \end{bsmatrix}}$.
Panel (a) shows the empirical coverage for 1000 replications as function of the sample size for different bootstrap methods (BS = nonparametric; PB = parametric; WLB = weighted likelihood bootstrap with weights generated from a Dirichlet with different values of $\alpha$) and two type of bootstrap intervals (perc = percentile; cperc = centered percentile).
Panel (b) show the ratio of bias (in absolute value) over the standard error for each bootstrap procedure as function of sample size.}
\label{fig:BivGaussian}
\end{figure}

\begin{table}
\caption{Simulation results for data generated from 10-dimensional independent $\chi^2$ distributions with $\text{df}=5$ degrees of freedom each. For each bootstrap method and sample size, the table reports averaged values of the entropy estimates, bootstrap estimates of bias, standard error, and percentile interval limits, along with their empirical coverage.}
\label{tab:ChiSq_df=5_d=10}
\small
\centering
\addtolength{\tabcolsep}{-0.6ex}
\begin{tabular}{lccccccccc}
\toprule
 & & & & \multicolumn{3}{c}{95\% percentile interval} & \multicolumn{3}{c}{95\% centered perc. interval} \\
\cmidrule(l{3pt}r{3pt}){5-7} \cmidrule(l{3pt}r{3pt}){8-10}
Sample size & \textbf{Estimate} & \textbf{Bias} & \textbf{SE} & lower & upper & \textbf{Coverage} & lower & upper & \textbf{Coverage}\\
\midrule
\multicolumn{10}{l}{Nonparametric bootstrap}\\[1ex]
 100 & 24.1444 & 0.0819 & 0.2494 & 23.5687 & 24.5447 & 0.865 & 23.7527 & 24.7167 & 0.971\\
 200 & 24.2166 & 0.0406 & 0.1719 & 23.8388 & 24.5103 & 0.940 & 23.9262 & 24.5912 & 0.914\\
 500 & 24.2210 & 0.0154 & 0.1103 & 23.9901 & 24.4206 & 0.925 & 24.0227 & 24.4534 & 0.937\\
1000 & 24.2423 & 0.0073 & 0.0778 & 24.0830 & 24.3868 & 0.913 & 24.0986 & 24.4027 & 0.897\\
\midrule
\multicolumn{10}{l}{Parametric bootstrap}\\[1ex]
 100 & 24.1444 & 0.0767 & 0.2575 & 23.5638 & 24.5660 & 0.868 & 23.7192 & 24.7280 & 0.985\\
 200 & 24.2166 & 0.0344 & 0.1800 & 23.8258 & 24.5328 & 0.954 & 23.8997 & 24.6038 & 0.970\\
 500 & 24.2210 & 0.0110 & 0.1133 & 23.9881 & 24.4290 & 0.953 & 24.0106 & 24.4513 & 0.937\\
1000 & 24.2423 & 0.0041 & 0.0800 & 24.0827 & 24.3945 & 0.926 & 24.0901 & 24.4028 & 0.912\\
\midrule
\multicolumn{10}{l}{Weighted likelihood bootstrap ($\alpha = 1$)}\\[1ex]
 100 & 24.1444 & 0.0804 & 0.2448 & 23.5838 & 24.5416 & 0.864 & 23.7513 & 24.7028 & 0.969\\
 200 & 24.2166 & 0.0400 & 0.1693 & 23.8457 & 24.5068 & 0.939 & 23.9232 & 24.5859 & 0.927\\
 500 & 24.2210 & 0.0150 & 0.1105 & 23.9909 & 24.4222 & 0.923 & 24.0211 & 24.4495 & 0.951\\
1000 & 24.2423 & 0.0073 & 0.0777 & 24.0845 & 24.3877 & 0.913 & 24.0973 & 24.4004 & 0.897\\
\midrule
\multicolumn{10}{l}{Weighted likelihood bootstrap ($\alpha = 4$)}\\[1ex]
 100 & 24.1444 & 0.0216 & 0.1233 & 23.8821 & 24.3635 & 0.542 & 23.9248 & 24.4058 & 0.588\\
 200 & 24.2166 & 0.0095 & 0.0852 & 24.0422 & 24.3745 & 0.728 & 24.0592 & 24.3934 & 0.700\\
 500 & 24.2210 & 0.0038 & 0.0552 & 24.1100 & 24.3258 & 0.615 & 24.1166 & 24.3320 & 0.628\\
1000 & 24.2423 & 0.0016 & 0.0390 & 24.1650 & 24.3175 & 0.547 & 24.1674 & 24.3197 & 0.564\\
\midrule
\multicolumn{10}{l}{Weighted likelihood bootstrap ($\alpha = 0.8137$)}\\
 100 & 24.1444 & 0.0990 & 0.2690 & 23.5188 & 24.5698 & 0.894 & 23.7160 & 24.7718 & 0.985\\
 200 & 24.2166 & 0.0471 & 0.1879 & 23.8032 & 24.5396 & 0.955 & 23.8983 & 24.6327 & 0.956\\
 500 & 24.2210 & 0.0179 & 0.1216 & 23.9673 & 24.4413 & 0.953 & 24.0003 & 24.4758 & 0.954\\
1000 & 24.2423 & 0.0084 & 0.0865 & 24.0662 & 24.4029 & 0.955 & 24.0830 & 24.4170 & 0.941\\
\bottomrule
\end{tabular}
\end{table}

\begin{figure}
\centering
\includegraphics[width=\textwidth]{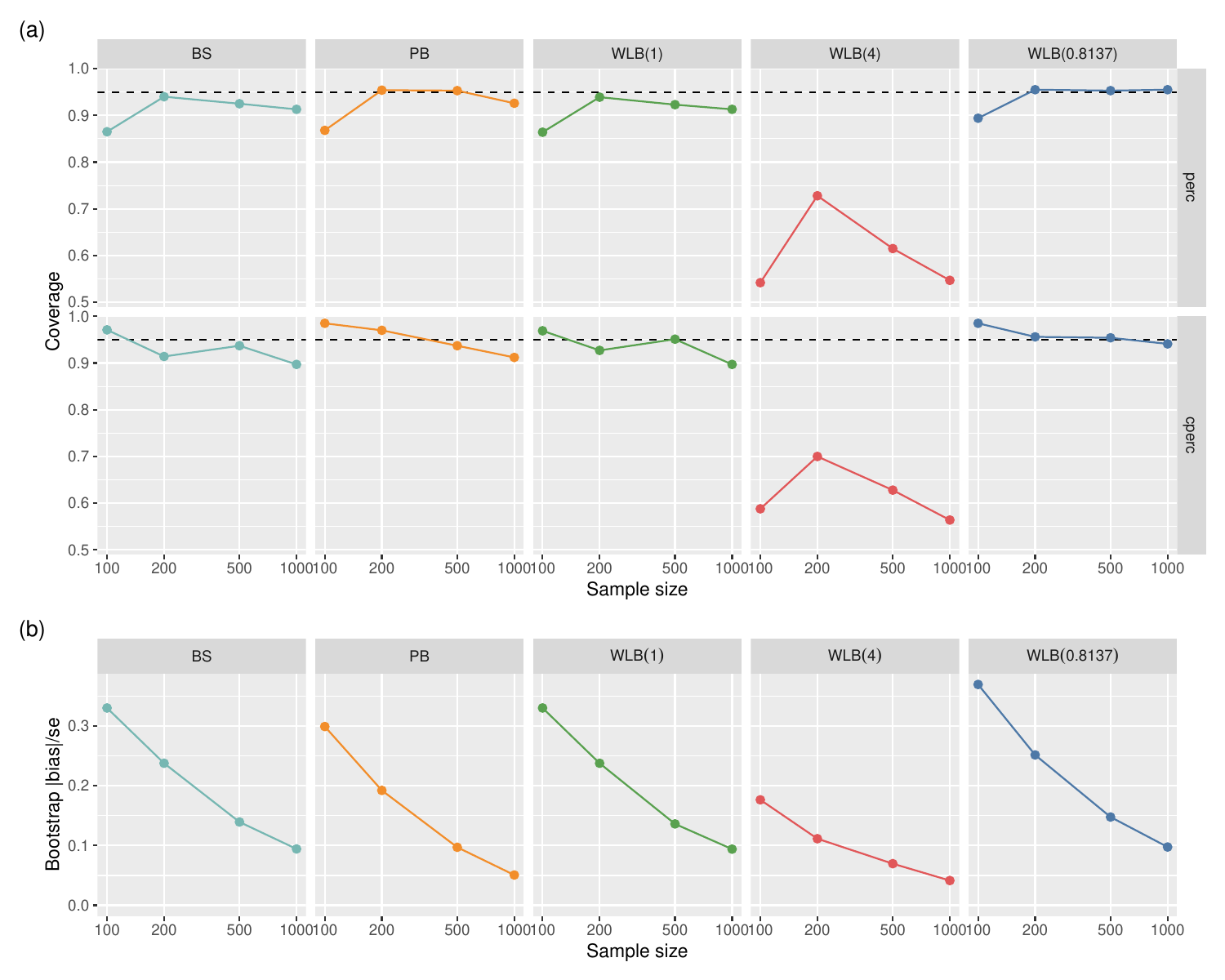}
\caption{Simulation results for data generated from 10-dimensional independent $\chi^2$ distributions with $\text{df}=5$ degrees of freedom each.
Panel (a) shows the empirical coverage for 1000 replications as function of the sample size for different bootstrap methods (BS = nonparametric; PB = parametric; WLB = weighted likelihood bootstrap with weights generated from a Dirichlet with different values of $\alpha$) and two type of bootstrap intervals (perc = percentile; cperc = centered percentile).
Panel (b) show the ratio of bias (in absolute value) over the standard error for each bootstrap procedure as function of sample size.}
\label{fig:ChiSq_df=5_d=10}
\end{figure}

\end{document}